\documentclass[10pt]{article}
\ifdefined\XeTeXversion\else
\usepackage[utf8]{inputenc}
\usepackage[T1]{fontenc}
\fi
\usepackage[a4paper,margin=2.2cm]{geometry}
\usepackage{amsmath,amssymb,amsfonts,amsthm}
\usepackage{graphicx,booktabs,multirow,array,enumitem,microtype,xcolor}
\usepackage[numbers,sort&compress]{natbib}
\usepackage{hyperref}
\usepackage[section]{placeins}
\usepackage{caption,longtable}
\hypersetup{colorlinks=true,linkcolor=black,citecolor=blue,urlcolor=blue,
pdftitle={Construction--Reuse Trade-offs for Exact Certificates in Fixed-Rank Threshold Screening},
pdfauthor={Zisu Li; Jingyang Du}}
\title{Construction--Reuse Trade-offs for Exact Certificates\\in Fixed-Rank Threshold Screening}
\author{%
Zisu Li$^{1}$ and Jingyang Du$^{2}$\\[0.8em]
\small $^{1}$School of Mechanical and Electrical Engineering, Central South University,\\
\small Changsha 410083, China\\
\small $^{2}$School of Journalism and Communication, Minzu University of China,\\
\small Beijing 100081, China\\
\smallskip
}
\date{}
\begin{document}
\maketitle
\section*{ABSTRACT}
Repeated threshold queries may reuse selected identities without reusing stale reports, but cheaper certificates need not shorten the complete response. We study selected-lower, atomic-upper (SLA) certificates for fixed-rank conjunctive screening with explicit missing-information semantics. An endpoint characterization and a counterexample separate same-source containment from policy-dependent online behavior. The original 320-session experiment reduces summed construction medians by 31.18\% against an exclusion-cover certificate, yet its Cover/SLA full-API geometric time ratio is 0.9682 (95\% conditional blocked interval 0.9593--0.9769), and SLA takes 12.94\% more summed time than uncached Bitmap. Three separately launched complete repeats preserve this adverse ordering, with Cover/SLA ratios of 0.9665--0.9718. An additional 720-configuration exploration varies catalogue size, construction period, requested count and query locality on empirically resampled tables. SLA is faster in 295 configurations against Cover and 271 against Bitmap, descriptive counts that do not establish universal superiority. Separately instrumented additive costs distinguish construction savings from retrieval and report costs. A plane-stress component case adds an independent analytical displacement check and 4,608 boundary-challenging queries: five implementations agree exactly, while medium- and fine-mesh selections differ at 459 positions. A state-stratified public bolt-record exercise preserves 185 incomplete positions among 1,479 requests. Raw timings, complete configuration results and a tested clean-environment package support reproducibility. The SLA construction was explored and refined through the self-evolving AI system ZiYor; the named authors specified, implemented and evaluated it. This is a bounded mechanics-to-query study, not physical joint qualification or universal speedup.

\noindent\textbf{Keywords:} Fixed-rank screening; Orthant top-k queries; Exact result reuse; Certificate construction; Engineering data tables; Controlled benchmarking

% Auto-generated from canonical JSON; see conversion_audit.json.

\section*{Introduction}
\label{sec:1}

Engineering selection frequently involves screening an existing set of records against changing requirements. A designer may revise bounds on dimensions, resource use or operating conditions while retaining a fixed preference order. The computational subproblem is then to return the first k records satisfying all stated predicates. This subproblem can occur inside a component-selection workflow, but it is not the entire workflow: geometric compatibility, physical capacity, assembly access and regulatory compliance require their own valid models and evidence. Efficient screening is useful only if its outputs preserve those distinctions.

A repeated query need not always repeat candidate retrieval. If the previous answer remains valid over a region of threshold space, a stored certificate can justify reusing its selected identities. Such reuse is well established in range-query, cached-view and safe-region research \cite{ref1,ref2,ref3,ref4,ref5}. The practical question examined here is narrower than whether caching works in principle. When candidate retrieval already uses an efficient bitmap backend, does reducing certificate-construction work improve the complete screening operation, or does the accompanying reduction in reuse opportunities offset the saving?

Two measurement choices make this question nontrivial. First, comparing certificates built at different queries conflates geometric differences with the evolving cache state. A region can be wider at an identical source query without being the region retained by an online method later. Second, returning a cached identifier list is less work than producing a current report. Selected records may remain unchanged while their margins relative to the new thresholds change. Timing only the certificate-membership check omits this necessary work and can misrepresent the benefit available to a downstream selection tool.

We examine selected-lower, atomic-upper certificates, abbreviated SLA throughout this article; the abbreviation does not denote a service-level agreement. SLA replaces the lower endpoints of an atomic threshold cell with the coordinate-wise maxima of the selected records while keeping the atomic upper endpoints. Its construction avoids an explicit scan of the competing prefix. We compare this certificate with the atomic cell, a greedily constructed exclusion-cover certificate, an uncached bitmap implementation and an uncached NumPy implementation. The three certificate methods use the same nonaccumulating construction-permission policy and the same full response interface. The candidate construction was explored and refined through the self-evolving AI system ZiYor; the named authors specified, implemented and evaluated the resulting method.

We separate four quantities: construction time at identical source queries, geometric coverage of a fixed future stream, realized online hits under a common permission policy, and complete-response time. The contribution is an explicit fixed-rank contract and a reproducible empirical comparison, rather than a new general caching principle. Besides preserving the original adverse primary outcome, we add three complete fresh-process repeats, a full size/period/count/locality matrix, additive cost diagnostics, and boundary-focused engineering checks. The resulting evidence explains where a local saving fails to become an end-to-end improvement and which limited workload conditions change the ordering.

\section*{Related work and positioning}
\label{sec:2}

Fixed-rank threshold screening is a special case of orthogonal range top-k retrieval on weighted points. Rahul et al. \cite{ref4} study top-k queries for orthogonal ranges with fixed weights. Their treatment also recognizes answer invariance between neighboring input coordinates. Consequently, neither the underlying query problem nor the existence of atomic threshold cells is introduced here. Our certificate layer follows an exact retrieval backend and does not claim a stronger range-search data-structure bound.

Spatial safe-region work explicitly connects region construction with subsequent reuse \cite{ref1,ref2}. Huang et al. \cite[Sect. 2.1, Definitions 1--3; Sects. 3.2 and 4.3--4.4]{ref2} hold keywords and k fixed while the query location changes; cached traversal information and conservative polygonal regions support client-side membership tests. Xie et al. \cite[Sect. 3.2, Algorithm 2; Sects. 3.3 and 4.1]{ref5} answer changing scoring queries from top-k views using LP thresholds, basis reuse and certain answers. McClain et al. \cite[Sects. 3.1--3.4, Algorithm 1; Sect. 4.3]{ref3} cache bitmap query segments and combine cached and remainder vectors, using CLOCK replacement. These are distinct reuse objects and query changes, not inferior implementations of the present interface.

Keller and Basu \cite[Sects. 3.1--3.2, Definitions 1--4; Sects. 4 and 5.1--5.5]{ref7} distinguish conservative client and liberal server descriptions, update notification, completeness and reclamation. Dar et al. \cite[Sects. 2.4 and 3.1--3.4]{ref8} use semantic regions, probe/remainder splitting and region replacement; their simulation excludes updates (Sect. 4.2). Thus conservative descriptions, reusable empty answers and cache maintenance are not new claims. The later geometric treatment also corrects earlier dynamic bounds \cite{ref9}; no improved dynamic bound is claimed here. Table~\ref{tab:main-1} condenses the distinctions; Supplement S1 provides all six comparison axes and page/algorithm locators.

\begingroup
\small
\setlength{\tabcolsep}{4pt}
\begin{longtable}{@{}>{\raggedright\arraybackslash}p{0.175182\dimexpr\textwidth-6\tabcolsep\relax}>{\raggedright\arraybackslash}p{0.364964\dimexpr\textwidth-6\tabcolsep\relax}>{\raggedright\arraybackslash}p{0.459854\dimexpr\textwidth-6\tabcolsep\relax}@{}}
\caption{Closest-method distinctions; detailed six-axis comparison in Supplement S1.}\label{tab:main-1}\\
\toprule
\textbf{Work} & \textbf{Reuse object / changing query} & \textbf{Distinction from this study} \\
\midrule
\endfirsthead
\multicolumn{3}{l}{\small\itshape Table \thetable{} (continued)}\\
\toprule
\textbf{Work} & \textbf{Reuse object / changing query} & \textbf{Distinction from this study} \\
\midrule
\endhead
\midrule
\multicolumn{3}{r}{\small\itshape Continued on next page}\\
\endfoot
\bottomrule
\endlastfoot
Huang et al. \cite{ref2} & Spatial answer and safe region / location & Fixed rank and hard thresholds here; not spatial-relevance ranking \\
McClain et al. \cite{ref3} & Partial bitmap vectors / query segments & Identity certificate here; current report still rebuilt \\
Xie et al. \cite{ref5} & Top-k views / scoring and k & Accessible fixed catalogue here; UNKNOWN is field-level \\
Keller--Basu \cite{ref7}; Dar et al. \cite{ref8} & Predicate/semantic regions and tuples / predicates & Single certificate and explicit construction permissions here \\
Ehlers \cite{ref6} & Top-k semantic caching; full method not inspected & Direct comparison gap; distinctness not established \\
\end{longtable}
\endgroup

The baselines are controlled implementations under one response contract, not reproductions of the cited complete systems. Published latencies are not imported into our hardware comparison. Ehlers's directly relevant thesis \cite{ref6} is identified through institutional metadata, but the full text remains inaccessible for method-level inspection; we cannot exclude overlap with its certificate rules. This limitation prevents exhaustive novelty clearance and is not repaired by adding more metadata citations.

\section*{Screening contract and exact certificates}
\label{sec:3}

\subsection*{Fixed ranking, completeness and current reports}
\label{sec:3-1}

Let a catalogue contain N records, sorted once by an immutable strict key consisting of a scalar score and the original row identifier. Smaller keys are preferred. We use zero-based rank positions i in the mathematical definitions, while exposed row identifiers remain one-based original data positions. Each record has d features. A supported query specifies a finite threshold vector t and a positive integer k. The permitted predicates are conjunctive upper bounds; neither the score nor the predicate family changes within a cache epoch.

A known feature either passes its bound or fails it. A record with any known failure is excluded, even if another field is missing. A record with no known failure but at least one missing field is UNKNOWN, not feasible. The known feasible set is defined by Eq. (1), and S is its first k records in the fixed order, or all its records when fewer than k are available. No imputation is used.

\begin{equation}
F(t)=\{i:x_i\ \mathrm{fully\ known},\ x_{ij}\leq t_j\ \forall j\}
\tag{1}\label{eq:main-1}
\end{equation}

Completeness is separate from feasibility of the returned records. If S contains k records, an unexcluded UNKNOWN record before the last selected rank can change the answer and blocks completeness. If fewer than k are selected, every unexcluded UNKNOWN record can matter. A complete result additionally requires the catalogue to be declared complete. That declaration concerns candidate coverage and does not imply that every feature is known. Define the competing prefix and its unselected records as in Eq. (2); for a complete result, each member of U has at least one known failing coordinate.

\begin{equation}
p=\begin{cases}\max(S)+1,& |S|=k,\\N,& |S|<k,\end{cases}\qquad U=\{0,\ldots,p-1\}\setminus S
\tag{2}\label{eq:main-2}
\end{equation}

Only complete results are eligible for a certificate. An incomplete query still returns its known feasible records and unresolved identifiers, but it cannot create a certificate asserting that its answer is complete. Changing the catalogue, ranking rules or k starts a new epoch and invalidates stored state. Unsupported queries do not advance the cache. Invalid inputs are rejected before any state read or modification.

Every successful supported call creates a new nested report: current thresholds and k, ordered selected row identifiers, original selected features and scores, current margins t minus x, unresolved identifiers, completeness and answer status. Metadata such as hit count may differ between methods; these operational counters are excluded from semantic equality. Reuse concerns selected identities, not a stale report object. Previously returned nested objects cannot be mutated to corrupt later reports. Table~\ref{tab:main-2} summarizes the notation and its computational meaning.

\begingroup
\small
\setlength{\tabcolsep}{4pt}
\begin{longtable}{@{}>{\raggedright\arraybackslash}p{0.136000\dimexpr\textwidth-6\tabcolsep\relax}>{\raggedright\arraybackslash}p{0.408000\dimexpr\textwidth-6\tabcolsep\relax}>{\raggedright\arraybackslash}p{0.456000\dimexpr\textwidth-6\tabcolsep\relax}@{}}
\caption{Notation and interpretation.}\label{tab:main-2}\\
\toprule
\textbf{Symbol} & \textbf{Definition} & \textbf{Qualification} \\
\midrule
\endfirsthead
\multicolumn{3}{l}{\small\itshape Table \thetable{} (continued)}\\
\toprule
\textbf{Symbol} & \textbf{Definition} & \textbf{Qualification} \\
\midrule
\endhead
\midrule
\multicolumn{3}{r}{\small\itshape Continued on next page}\\
\endfoot
\bottomrule
\endlastfoot
N, d, k & Catalogue size, dimension, requested count & Ranking and catalogue fixed within an epoch \\
x, t & Original features and current upper thresholds & Native units per coordinate \\
S, U & Selected ranks and competing unselected prefix & A complete result is required for certification \\
\ensuremath{V_j}, \ensuremath{a_j}, \ensuremath{b_j} & Sorted unique feature values and adjacent endpoints & Missing fields use an internal sentinel only \\
\ensuremath{\ell_j} & Maximum selected feature in coordinate j & \ensuremath{-}\ensuremath{\infty} when S is empty \\
\ensuremath{C_A}, \ensuremath{C_L}, \ensuremath{C_G} & Atomic, SLA and exclusion-cover boxes & Half-open upper boundaries \\
\ensuremath{U_{\max}}, w & Largest unique-value count; bitmap storage words & Bitmap arithmetic is not constant-time \\
T, R & Session-median total and baseline/target ratio & R > 1 favors the target \\
\end{longtable}
\endgroup

\begin{figure}[!htbp]
\centering
\includegraphics[width=\textwidth,height=.64\textheight,keepaspectratio]{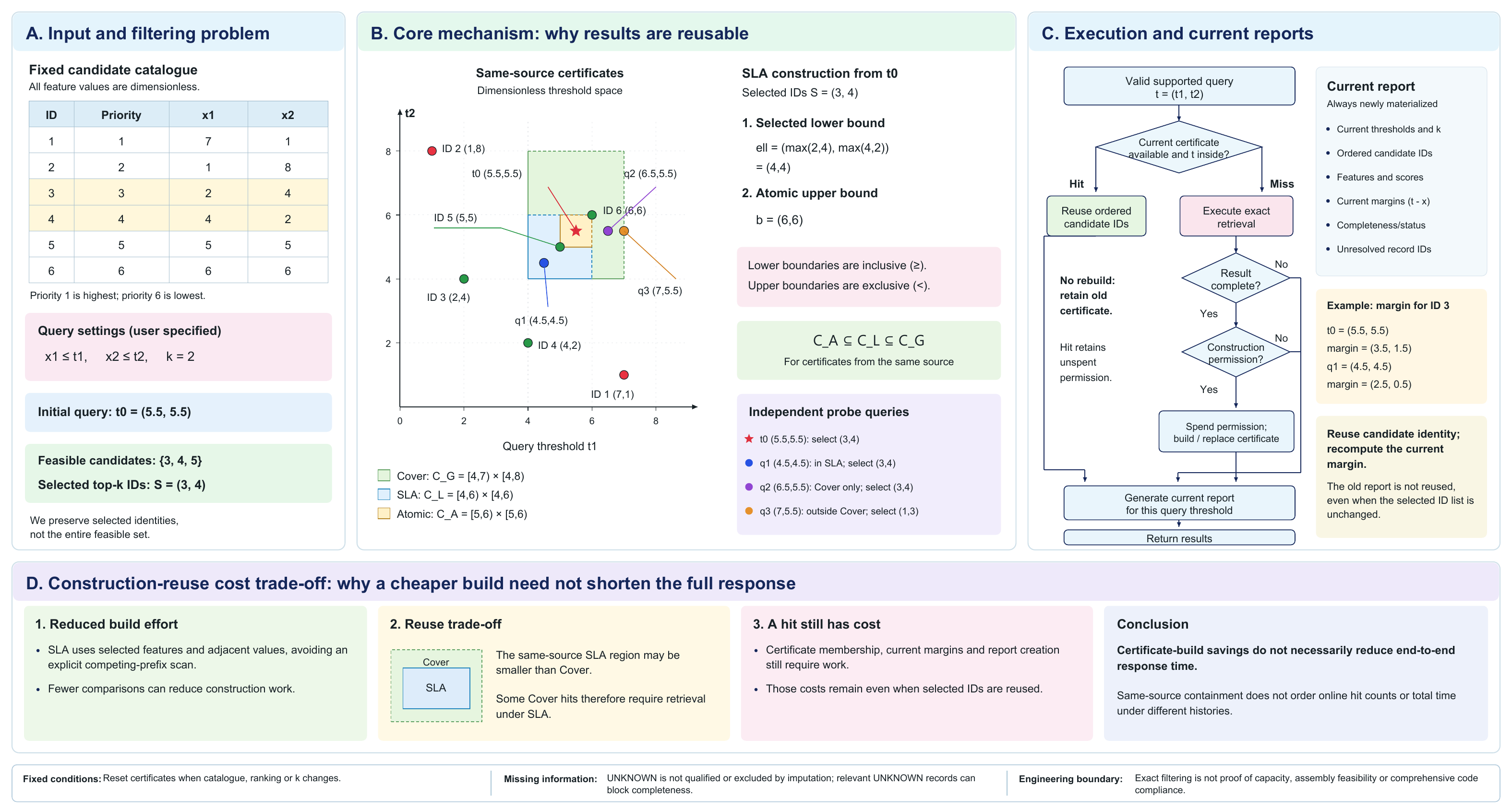}
\caption{Selected-lower, atomic-upper (SLA) certificates: construction, reuse and cost trade-offs. This is an illustrative dimensionless six-record example, not a benchmark measurement. At $t_0=(5.5,5.5)$ and $k=2$, the selected IDs are $(3,4)$; $C_A=[5,6)\times[5,6)$, $C_L=[4,6)\times[4,6)$ and $C_G=[4,7)\times[4,8)$. The illustrated queries are independent probes of the certificates built at $t_0$, not a sequential online trajectory. Reuse preserves identities, while each call constructs a current report. Construction requires both a complete miss and available permission. Same-source containment does not imply online hit or time dominance, and the diagram does not establish physical connection qualification.}
\label{fig:1}
\end{figure}

Figure~\ref{fig:1} illustrates the certificate mechanism using a dimensionless example; its numbers are explanatory and are not additional benchmark observations.

\subsection*{SLA construction and correctness}
\label{sec:3-2}

For each coordinate, an immutable sorted unique-value table \ensuremath{V_j} supports predecessor and successor searches. The comparison representation includes an internal negative-infinity sentinel for missing fields so that missingness cannot falsely exclude a record. This sentinel is not a physical measurement. Let \ensuremath{a_j} and \ensuremath{b_j} be the adjacent endpoints around the source threshold, as in Eq. (3); maxima and minima of the empty sets in that equation use negative and positive infinity, respectively. The upper endpoint is strictly greater than the source threshold.

\begin{equation}
\begin{aligned}a_j&=\max\{v\in V_j:v\leq t_j\},\\b_j&=\min\{v\in V_j:v>t_j\}\end{aligned}
\tag{3}\label{eq:main-3}
\end{equation}

The atomic certificate preserves the status of every coordinate comparison inside its cell. SLA relaxes only its lower endpoints. Let \ensuremath{\ell_j} be the maximum value of coordinate j among selected records, with negative infinity for an empty selection, as in Eq. (4). The two half-open boxes are given in Eq. (5). Both contain the source query when that query has a complete result.

\begin{equation}
\ell_j=\begin{cases}\max_{i\in S}x_{ij},&S\ne\varnothing ,\\-\infty,&S=\varnothing ,\end{cases}
\tag{4}\label{eq:main-4}
\end{equation}

\begin{equation}
\begin{aligned}C_A&=\prod_{j=1}^{d}[a_j,b_j),\\C_L&=\prod_{j=1}^{d}[\ell_j,b_j)\end{aligned}
\tag{5}\label{eq:main-5}
\end{equation}

Proposition 1 (exact identity reuse). Under the fixed-epoch contract, every valid threshold vector in \ensuremath{C_L} has the same ordered selected records and completeness as its complete source result. To see this, every selected record remains feasible because each of its coordinates is at most \ensuremath{\ell_j}, which is at most the new threshold. Each competing record originally had a known failing coordinate j. Its feature value is strictly above the source threshold, hence is at least \ensuremath{b_j}. The new threshold is strictly below \ensuremath{b_j}, so that failure remains. With k selected records, later records cannot affect the first k; with fewer than k, the competing prefix includes every remaining record. The argument also covers an empty selection. An UNKNOWN competitor cannot be silently admitted because the source certificate was permitted only when every relevant unselected record had a known failure.

The proposition concerns ordered identities and completeness, not unchanged margins. A hit must still materialize the current report. The closed lower and open upper boundaries are essential: a selected record remains feasible at its largest coordinate value, whereas an excluded record can become feasible at an upper endpoint. Degenerate catalogues and k exceeding N do not require changing this convention.

Proposition 2 (restricted lower expansion). Fix the source atomic upper vector b and consider nonempty half-open boxes with lower vector L that contain the source query. Over finite real threshold vectors, every point of such a box preserves feasibility of S if and only if \ensuremath{L_j} is at least \ensuremath{\ell_j} for every coordinate. Sufficiency follows coordinate by coordinate. For necessity, if some \ensuremath{L_j} is below \ensuremath{\ell_j}, choose that threshold strictly between them and keep all other thresholds at their source values. A selected record attaining \ensuremath{\ell_j} then fails. Therefore \ensuremath{C_L} is the largest box in this fixed-upper class that preserves S. This direct characterization is not a globally maximal safe region: changing the upper endpoints can yield a larger certificate, and imposing a restricted physical query domain changes the relevant set intersections.

\subsection*{Exclusion cover, complexity and the digital domain}
\label{sec:3-3}

The exclusion-cover certificate uses the same selected lower vector but obtains upper bounds from an assignment of competitors to known failing coordinates. Each competitor in U is assigned to one coordinate where it fails at the source query. The implementation greedily chooses the coordinate excluding the largest number of still-unassigned competitors, breaking ties by the lowest coordinate index. For the assigned subset \ensuremath{A_j}, the upper endpoint is its minimum feature value, or positive infinity if that subset is empty. Equation (6) defines the resulting box. Every assigned value exceeds the source threshold, and therefore is at least its atomic successor.

\begin{equation}
\begin{aligned}u_j&=\min_{i\in A_j}x_{ij}\quad(A_j\ne\varnothing ),\\u_j&=+\infty\quad(A_j=\varnothing ),\\C_G&=\prod_{j=1}^{d}[\ell_j,u_j)\end{aligned}
\tag{6}\label{eq:main-6}
\end{equation}

For an identical source query, selected set, catalogue and index, \ensuremath{\ell_j} is at most \ensuremath{a_j} and \ensuremath{u_j} is at least \ensuremath{b_j}. The containment relation in Eq. (7) follows. It compares certificates constructed at one source; it does not compare the different certificates an online policy may retain after different histories. Region coverage below is measured by membership of fixed subsequent queries, not by a volume combining incompatible units or infinite endpoints.

\begin{equation}
C_A\subseteq C_L\subseteq C_G
\tag{7}\label{eq:main-7}
\end{equation}

After the immutable column indices have been prepared, SLA needs adjacent-value searches and a scan of the selected features, rather than a competing-prefix scan. Its construction cost is bounded by the first expression in Eq. (8). The cover implementation uses bigint bitmaps for greedy assignment and binary searches for assigned minima; a worst-case bound is given alongside it. Here \ensuremath{U_{\max}} is the largest column unique-value count and w is the number of bitmap storage words, which grows with N. Bigint intersection and population count are not constant-time operations. Certificate storage is O(d + |S|). These bounds exclude catalogue preparation, retrieval on misses and report construction.

\begin{equation}
\begin{aligned}\mathrm{SLA}:&\quad O(d(1+\log(1+U_{\max}))+d|S|),\\\mathrm{Cover}:&\quad O(d^2w+dw\log(1+U_{\max})+d|S|)\end{aligned}
\tag{8}\label{eq:main-8}
\end{equation}

A certificate hit still checks d bounds and creates selected-record fields and margins, requiring work that depends on d and |S|. It is therefore not a constant-time full response. The bitmap backend intersects per-coordinate candidate masks and stops early only when the entire candidate mask becomes empty. An empty known-feasible subset alone is not a valid early exit because UNKNOWN records can still affect completeness. A vectorized NumPy backend provides a separate uncached comparator within the same interface.

The real-valued proof and the implemented numeric domain are distinguished. The implementation accepts finite binary64 values and integers exactly representable as binary64, rejects booleans as numeric inputs and does not silently round larger integers. A shared pre-state check uses immutable column minima over fully known records to reject thresholds that could produce a nonfinite positive margin. This conservative report-representability condition can reject inputs even when the offending record would not be selected. Implemented boxes are interpreted as intersections with this admissible digital domain; Proposition 2 is not a claim of unique endpoints on a discrete domain. Invalid-input, boundary, missing-field, epoch, fallback and output-mutation cases were exercised separately from the public-data timing experiment.

\subsection*{Equal construction permissions and a counterexample}
\label{sec:3-4}

A nonaccumulating construction permission is made available at positions 1, 1+P, 1+2P, \ldots{} of a valid-query session. A hit retains it; an eligible complete miss spends it on construction and replaces the certificate on success. A miss without permission retrieves the correct report while retaining the old certificate. Invalid/unsupported requests do not advance this schedule. The original matrix fixes P = 32 (at most four attempts in 128 queries); the supplement varies P. Equal permissions do not imply equal realized builds or source positions.

The following deterministic example shows why Eq. (7) alone cannot establish online hit dominance. Consider three fully known one-dimensional records (identifier, feature, score): (1, 2, 0), (2, 0, 1) and (3, 1, 2), with k = 1 and smaller scores preferred. Queries 1--32 have threshold 0.5; query 33 has 1.5; query 34 has 2.5; and queries 35--64 have 1.5. Table~\ref{tab:main-3} follows the actual permission policy. All selected identities are record 2 except at query 34, where record 1 is selected.

At the first query, SLA stores [0, 1), while Cover stores [0, 2). Both hit the next 31 queries. At query 33, SLA spends its replenished token on [0, 2), whereas Cover hits and retains its token. At query 34, SLA cannot rebuild and keeps its old box; Cover replaces its box with [2, +\ensuremath{\infty}). The last 30 queries consequently hit SLA's retained box but miss Cover's new one, with no token available to replace it. SLA records 61 hits and three misses, compared with Cover's 32 hits and 32 misses, despite two builds for each. Separate execution checked all 128 reports from the two methods against the scalar reference. The example disproves a containment-to-online-hit-dominance implication under this policy; it neither compares elapsed time nor proves universal two-way nondominance or an optimal policy.

\begingroup
\small
\setlength{\tabcolsep}{4pt}
\begin{longtable}{@{}>{\raggedright\arraybackslash}p{0.250000\dimexpr\textwidth-8\tabcolsep\relax}>{\raggedright\arraybackslash}p{0.250000\dimexpr\textwidth-8\tabcolsep\relax}>{\raggedright\arraybackslash}p{0.250000\dimexpr\textwidth-8\tabcolsep\relax}>{\raggedright\arraybackslash}p{0.250000\dimexpr\textwidth-8\tabcolsep\relax}@{}}
\caption{One-based query trace for the deterministic cache-state counterexample.}\label{tab:main-3}\\
\toprule
\textbf{Queries} & \textbf{Threshold} & \textbf{SLA state/action} & \textbf{Cover state/action} \\
\midrule
\endfirsthead
\multicolumn{4}{l}{\small\itshape Table \thetable{} (continued)}\\
\toprule
\textbf{Queries} & \textbf{Threshold} & \textbf{SLA state/action} & \textbf{Cover state/action} \\
\midrule
\endhead
\midrule
\multicolumn{4}{r}{\small\itshape Continued on next page}\\
\endfoot
\bottomrule
\endlastfoot
1 & 0.5 & Miss; build [0, 1) & Miss; build [0, 2) \\
2--32 & 0.5 & 31 hits & 31 hits \\
33 & 1.5 & Miss; build [0, 2) & Hit; retain token \\
34 & 2.5 & Miss; no token; retain [0, 2) & Miss; build [2, +\ensuremath{\infty}) \\
35--64 & 1.5 & 30 hits & 30 misses; no token \\
Total & 64 queries & 61 hits; 3 misses; 2 builds & 32 hits; 32 misses; 2 builds \\
\end{longtable}
\endgroup

\section*{Experimental design}
\label{sec:4}

\subsection*{Data and original matrix}
\label{sec:4-1}

The experiment uses the official UCI Airfoil Self-Noise \cite{ref10} and Concrete Compressive Strength \cite{ref11} archives. The airfoil data originate from the work associated with the NASA self-noise report \cite{ref12}; the concrete archive preserves the attribution to Yeh's strength-modeling study \cite{ref13}. These are observed engineering-related tables, not collected connector catalogues. Dataset choice preceded performance-query generation and was based on interpretable native fields and accessible original data. No dataset was removed after inspecting timing results.

Airfoil contains 1,503 records, five inputs and a measured sound-pressure response. The five inputs are used as upper-threshold features; the response supplies a fixed ascending preference score. Concrete contains 1,030 records, eight inputs and measured compressive strength. Its first seven inputs are ingredient quantities, followed by age. We negate measured strength for ascending ranking, so higher observed strength is preferred. Both rankings use the original one-based row identifier to resolve equal scores. These are within-catalogue rankings of measured responses, not predictions at unobserved conditions or estimates of physical design optimality.

Both original files are parsed at source precision, retaining all rows. Airfoil has no duplicate complete rows; Concrete has 25 additional identical full rows and 34 additional identical input rows, including nine equal-input groups with different measured responses. No missing/nonfinite values occur in these two tables. No averaging or train/test predictive modeling is performed. Native units and ranges are retained in Supplement S2. The original 32 groups cross two datasets, k \ensuremath{\in} \{1,5\}, broad/positive strata and iid/local/shuffled/jumps orders. Ten generator replicates per group give 320 sessions of 128 positions; ten technical repeats per method yield 2,048,000 timed calls.

The iid coordinates are uniform on [0,1]. Local coordinates start uniformly in [0.2,0.8] and undergo clipped normal increments of standard deviation 0.015; shuffled permutes exactly the same 128 local points. Jumps resets its center every 16 positions with 0.015 normal perturbations. With observed minima m and maxima M, broad thresholds use Eq. (9). Positive thresholds start at h, the coordinate-wise maximum of k rows sampled without replacement, guaranteeing at least k known feasible records. All four orders share an anchor; all queries and independent scalar-reference reports are frozen before timing. These artificial thresholds are not measurement noise or physically realizable new experiments. Supplement S2 gives the deterministic seed labels.

\begin{equation}
\begin{aligned}t_j^{\mathrm{broad}}&=m_j+u_j(M_j-m_j),\\t_j^{\mathrm{positive}}&=h_j+u_j(M_j-h_j)\end{aligned}
\tag{9}\label{eq:main-9}
\end{equation}

\subsection*{Timing boundaries, repeats and estimands}
\label{sec:4-2}

The same-origin experiment isolates certificate construction from different online histories. At fixed zero-based source positions 0, 32, 64 and 96 of every session, all three constructors receive the same complete result, thresholds and prepared index, regardless of whether an online method would build there. Constructor order is randomized, and each constructor runs ten times. Retrieval before construction and certificate validation afterward are outside this clock. This produces 1,280 source positions and 38,400 construction timings. A separate instrumented construction pass records diagnostics and is not mixed with the timed constructors.

Each source certificate is tested against its next 31 fixed queries, giving 39,680 potential coverage events per method. Online methods instead start fresh state and process all 128 calls. The full select clock includes request parsing, admissible-domain checking, state access, retrieval, any permitted construction and fresh nested report materialization. Initialization, explicit pre-repeat garbage collection, oracle checks and output destruction are excluded; automatic collection inside the call is included. Method order is randomized for every technical repetition. NumPy, Bitmap, Atomic, SLA and Cover return the same semantic fields, not just selected IDs.

The measurements use one CPU worker on a Windows 10 build-19045 host with two Intel Xeon Gold 6130 processors at 2.10 GHz, each with 16 physical cores and 32 logical processors. This is not a parallel speedup experiment. Python is 3.12.8, NumPy is 2.5.3 and the original XLS reader is xlrd 2.0.2. The performance counter has a reported resolution of 100 ns. No processor affinity, dedicated idle-host condition or explicit numerical-library thread limit was imposed. These conditions constrain the interpretation of small timing differences and cross-machine reproducibility.

Three additional complete runs are launched in separate processes on the same host with logical-CPU affinity \{0\} and OMP, OpenBLAS, MKL and NumExpr thread settings of one. Each repeats every original group, seed and technical repetition using Python 3.12.14 and NumPy 2.3.5. Before the technical repetitions, four untimed calls warm a separate instance of each method; every measured repetition still starts with a new instance. The screened core and oracle are unchanged, but these environment/warmup changes mean controlled repeats, not a byte-identical rerun of the original harness. Initial per-CPU load and per-session environment records are retained; other background jobs remain active, so no dedicated idle-host claim is made. All three completed runs are reported.

Let T be the median of repeated 128-call sums, Eq. (10), not the sum of per-query medians. Original-matrix runs use ten technical repeats. R in Eq. (11) weights the 32 groups equally and then their ten generator replicates; R > 1 favors SLA. The unique original primary contrast remains Cover/SLA. Each 95\% interval uses 10,000 paired blocked bootstrap draws: ten replicate clusters are sampled within each of eight dataset/k/stratum blocks, preserving all four query orders together \cite{ref14,ref15}. Thus each run has 320 sessions but 80 replicate clusters for resampling. Intervals are conditional on the measured run, tables and generators, not on an engineering-project population or pooled across runs. Eq. (12) separately measures signed change in summed session medians.

\begin{equation}
T_{m,s}=\operatorname{median}_{r=1,\ldots,10}\left(\sum_{q=1}^{128}\tau_{m,s,r,q}\right)
\tag{10}\label{eq:main-10}
\end{equation}

\begin{equation}
R=\exp\left[\frac{1}{32}\sum_{g=1}^{32}\frac{1}{10}\sum_{s\in g}\log\left(\frac{T_{\mathrm{base},s}}{T_{\mathrm{target},s}}\right)\right]
\tag{11}\label{eq:main-11}
\end{equation}

\begin{equation}
\Delta=100\left(\frac{\sum_sT_{\mathrm{target},s}}{\sum_sT_{\mathrm{base},s}}-1\right)\%
\tag{12}\label{eq:main-12}
\end{equation}

\subsection*{Full supplementary sensitivity and direct diagnostics}
\label{sec:4-3}

A frozen exploratory matrix crosses two source tables, N \ensuremath{\in} \{128,512,2048,8192,32768\}, P \ensuremath{\in} \{1,8,32,128\}, k \ensuremath{\in} \{1,5,20\}, broad/positive strata and iid/local015/local001 streams. Local increments have standard deviations 0.015 and 0.001; clipping and positive anchoring follow the original design. Its 720 configurations each use three resampling seeds, three technical repeats and 128 queries, giving 2,160 sessions and 4,147,200 full-API timings. Every row is sampled with replacement from the original empirical table and receives a distinct new identifier. N levels share pseudorandom draw prefixes within a seed. Empirical support therefore saturates at the original data; increasing N is not new physical diversity, and these are not independent held-out projects. All configurations and unfavourable outcomes are retained; no selected winner is presented as a confirmatory result.

For these sessions, Eq. (10) uses three rather than ten technical repeats. Configuration ratios are geometric means over the three catalogue/query seeds; figures pool explicitly stated cells descriptively without confidence or multiplicity claims. Compilation time is measured separately for each of the 30 dataset/size/seed catalogues. Retained Python objects belonging to column indices are counted separately from the feature array and whole-process RSS. Initial load and session preparation remain outside API clocks; the shared index is not an exclusive preprocessing cost of any one baseline.

A separate diagnostic pass times parse, domain, contains, retrieve, build and materialize at nonoverlapping top-level boundaries. Nested certificate checks inside construction stay inside build. The residual is the measured instrumented full call minus those six same-call stages, including timer and dispatch overhead; their integer sum equals every instrumented call exactly. These passes perturb runtime and are not subtracted from uninstrumented session medians. Across the three original-matrix repeats and sensitivity exploration, 10,291,200 main timings and 1,996,800 separately instrumented method/query positions are retained.

\subsection*{State-stratified public bolt records}
\label{sec:4-4}

A separate functional exercise uses the NIST double-shear bolt archive \cite{ref16} and its associated experimental description \cite{ref17}. Six original workbooks contain 92 internally identified specimens in 30 diameter-label, grade and temperature conditions, whereas the article describes 91 tests. The discrepancy is retained, not resolved by dropping a record. The archive force is total applied force across two shear planes, and displacement is measured through the apparatus rather than locally across a bolt. Fixture group 0 covers the 19/22 labels with the A36 apparatus; group 1 covers the nominal 25.4 mm bolts with a different fixture. These groups are queried separately, not treated as interchangeable connectors.

Source qualification precedes the workload. A 22/A325/200 \ensuremath{{}^\circ}C condition contains a diameter-label conflict between its worksheet name and internal identifier. A 25/A325/400 \ensuremath{{}^\circ}C condition has a failure-force summary of 50.06 kN where the article reports 500.6 kN. Both whole conditions retain known temperature and fixture labels but receive UNKNOWN for all four derived response features. This precaution does not prove that their curve ordinates are erroneous. We neither repair the source numerically nor discard conflicting rows. A single negative displacement point remains in the stored source and is excluded only by the nonnegative-window predicate.

The operational predicates are deliberately finite-record predicates. For bounds b = 0.5, 1, 2 and 4 mm, a specimen value is the maximum recorded total force among points with 0 \ensuremath{\leq} displacement \ensuremath{\leq} b up to and including its first global force maximum. The condition value Cb is the minimum of these specimen values across all archived repeats; an empty specimen window makes the condition value UNKNOWN. All 92 first maxima occur at their final published point, so the prefix convention actually retains all 4,370 published pairs. The four bounds are analyst-defined interrogation windows, not code deflection limits. No curve interpolation, fitted constitutive model, resistance factor or statistical lower tolerance bound is introduced. Different bounds may use different source points: satisfying four predicates does not certify a single simultaneous loading path.

Conditions are ranked lexicographically by integer diameter label, grade and temperature, with one-based row identifiers and row identifier as the scalar score. This is a fixed administrative preference proxy, not measured cost or a design optimum. We embed each condition as x = (\ensuremath{-}C0.5, \ensuremath{-}C1, \ensuremath{-}C2, \ensuremath{-}C4, T, \ensuremath{-}T, g, \ensuremath{-}g), and query with t = (\ensuremath{-}f0.5, \ensuremath{-}f1, \ensuremath{-}f2, \ensuremath{-}f4, T, \ensuremath{-}T, g, \ensuremath{-}g). Paired categorical coordinates enforce exactly the requested temperature T and fixture g. The force demands are finite, nonnegative and nondecreasing across the four windows. Supported temperatures are 20, 200, 400, 500 and 600 \ensuremath{{}^\circ}C, fixtures are 0/1, and the original workload uses k \ensuremath{\in} \{1,2,5\}. Neither temperature interpolation nor cross-fixture extrapolation is supported.

The original 5,460 positions retain their fixed wide-range workload (k = 1,2,5); they are not replaced by easier positive queries. The supplementary workload instead queries capacities actually attainable in each temperature/fixture condition: scaled known capacity vectors with factors 0, 0.5, 0.9, 1 and 1.1, and individual-coordinate equality/adjacent-binary64 probes that meet the nonnegative nondecreasing-demand rule, with k = 1,3,5. Invalid generated demand vectors are outside this declared query family, not failed results removed after execution. The retained 1,479 positions include duplicates from overlapping probes; there are 1,026 distinct requests. The same literal workbook-row reference and the original two quarantined conditions are retained. Two records each have four unknown response values: eight field values, not eight response-field types. Answered, empty and incomplete reports are tested separately for all five methods, without latency retiming or fitting a bolt model.

\subsection*{Verified elastic attributes and decision boundaries}
\label{sec:4-5}

A separate numerical case tests the path from a mechanics model to the screening interface. The publicly documented Kirsch plate benchmark supplies a reproducible elastic reference: E = 210,000 MPa, Poisson ratio 0.3, a 2,000 \ensuremath{\times} 2,000 mm square, hole radius a = 100 mm, thickness 100 mm and far-field tension \ensuremath{\sigma_{\infty}} = 0.001 MPa \cite{ref18}. The analytical elastic field is documented independently \cite{ref19}. For verification, we prescribe its exact traction \ensuremath{\sigma}\ensuremath{\cdot}n on both outer edges of the quarter-domain, not the uniform finite-boundary loading of the COMSOL example. This author-constructed exact-boundary problem removes outer-boundary truncation as a competing error source; it is not a reproduction of a commercial solver result. Along x = 0, the reference axial stress is given by Eq. (13), with \ensuremath{\sigma_{xx}}(0,a) = 3\ensuremath{\sigma_{\infty}}.

\begin{equation}
\sigma_{xx}(0,y)=\frac{\sigma_\infty}{2}\left[2+\left(\frac{a}{y}\right)^2+3\left(\frac{a}{y}\right)^4\right]
\tag{13}\label{eq:main-13}
\end{equation}

The implementation uses eight-node serendipity quadrilaterals, full 3 \ensuremath{\times} 3 Gauss integration and a true two-dimensional small-strain plane-stress constitutive law. For the strain vector (\ensuremath{\varepsilon_{xx}}, \ensuremath{\varepsilon_{yy}}, \ensuremath{\gamma_{xy}}), the normal block of C is E/(1\ensuremath{-}\ensuremath{\nu}\textsuperscript{2}) times [[1,\ensuremath{\nu}],[\ensuremath{\nu},1]], and the shear entry is E/[2(1+\ensuremath{\nu})]. Symmetry imposes \ensuremath{u_x} = 0 on x = 0 and \ensuremath{u_y} = 0 on y = 0; the circular hole is traction-free. No additional pin, artificial stiffness, contact law or out-of-plane degrees of freedom are introduced. Element assembly and the free-degree-of-freedom solve are defined by Eq. (14). Prescribed edge tractions are integrated consistently with five Gauss points. The sparse system is solved using SciPy spsolve with SuperLU \cite{ref20}.

\begin{equation}
\begin{aligned}\mathbf K_{ff}\mathbf u_f&=\mathbf f_f,\\\mathbf K_e&=t\int_{\Omega_e}\mathbf B^{T}\mathbf C\mathbf B\,dA\end{aligned}
\tag{14}\label{eq:main-14}
\end{equation}

The three reference meshes contain 192, 768 and 3,072 Q8 elements (641, 2,433 and 9,473 nodes). Positive Jacobians, hole stress at (0,a), and a 21-point stress profile y/a = 1,1.1,\ldots{},3 are checked. In addition, an independently transcribed analytical displacement field uses plane-stress \ensuremath{\kappa} = (3\ensuremath{-}\ensuremath{\nu})/(1+\ensuremath{\nu}), not a plane-strain coefficient \cite{ref21} (Eq. 3.106). Supplement S3 gives the polar formula and Cartesian conversion. Unweighted vector nodal relative L2 error is measured over all nodes and separately r \ensuremath{\leq} 3a, without translation, amplitude fitting or injection of exact displacements into the solve. Both finest-grid errors must be \ensuremath{\leq}1\%. Numerical differentiation plus the plane-stress constitutive law independently checks the formula against the analytical stresses.

\begin{figure}[!htbp]
\centering
\includegraphics[width=\textwidth,height=.64\textheight,keepaspectratio]{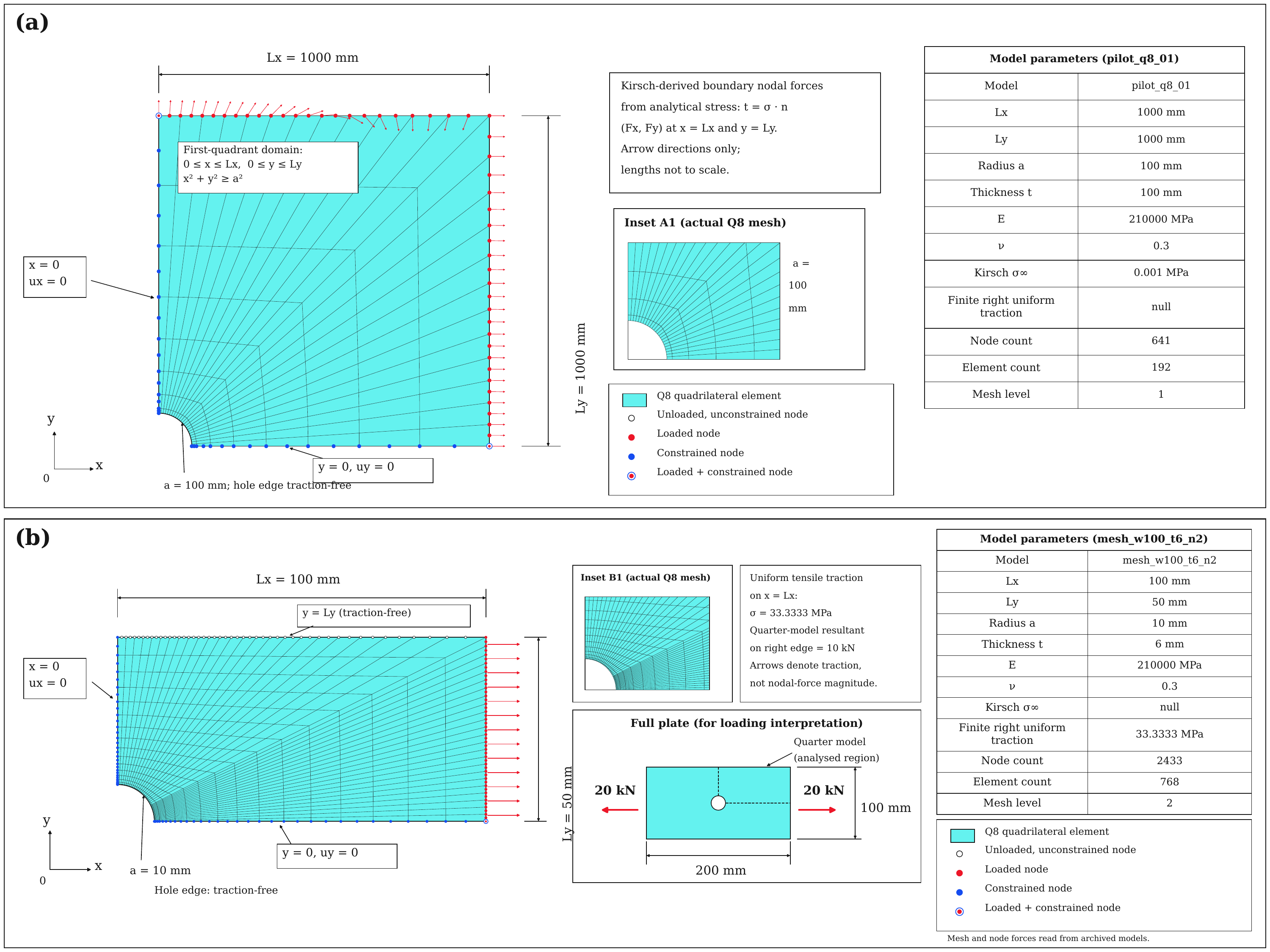}
\caption{Two distinct boundary-value problems: exact Kirsch tractions for verification (a) and uniform end loading for the finite catalogue (b). Symmetry and a traction-free hole apply to both; the finite plates are not loaded by bolt-hole contact.}
\label{fig:2}
\end{figure}

The downstream catalogue is a different, finite-boundary problem: full plate length 200 mm, hole radius 10 mm, widths 80, 100 and 120 mm, and thicknesses 4, 6 and 8 mm, giving nine candidates. Each full end carries 20 kN; the quarter-model right edge therefore carries 10 kN under uniform traction. Its top edge and hole remain unloaded. The material constants and symmetry conditions are unchanged. The two problems are contrasted in Figure~\ref{fig:2}. Each candidate is solved on the medium and fine meshes. Its attributes are hole \ensuremath{\sigma_{xx}}, the work-conjugate loaded-edge displacement \ensuremath{\bar{u}} = \ensuremath{\sum f_xu_x/\sum f_x}, minimum side ligament (W\ensuremath{-}2a)/2 and thickness. Here \ensuremath{\bar{u}} is the quarter-edge displacement; full end-to-end extension is 2\ensuremath{\bar{u}} by symmetry. Candidates are ranked by full net volume t(200W\ensuremath{-}\ensuremath{\pi}a\textsuperscript{2}), breaking ties by ID. These are analyst-defined elastic plates, not bolt-bearing or preloaded connections.

All nine finite candidates retain medium/fine solves, force/moment/energy residual checks and native-versus-reconstructed stresses; a central candidate has zero- and half-load checks. The original 49-query stress/displacement grid and 54 stored-value \ensuremath{\pm}10\textsuperscript{-9} probes remain. To challenge its narrow coverage, 108 new probes vary each of nine candidates' four stored attributes through the immediately lower binary64 value, equality and immediately higher value. Another 4,500 positions cross nine candidates, five fractions \{0,0.25,0.5,0.75,1\} between medium/fine stress values, the same five fractions for displacement, minimum ligaments \{20,30,40,50,51\} mm and minimum thicknesses \{4,6,8,9\} mm. These queries activate geometry constraints and include no-result cases; k remains one. The 4,608 positions contain 4,572 unique threshold vectors. All five complete reports are compared with independent scalar enumeration.

Two mesh-related outcomes are distinct. Direct disagreement compares the actual medium- and fine-mesh selected IDs, including the empty answer. Envelope ambiguity instead compares coordinate-wise optimistic and pessimistic attributes formed from those two meshes; combined extrema need not correspond to either actual mesh. This observed spread is neither a certified discretization-error bound nor a probability interval. Supplement S3 specifies the original mechanical tolerances and preserves the entire 23-solve matrix.

\section*{Results}
\label{sec:5}

\subsection*{Original outcome and three complete repeats}
\label{sec:5-1}

All 2,048,000 timed semantic reports agree with the independent scalar oracle, as do the separately checked diagnostics. Among the 40,960 original query positions, 35,250 are answered and 5,710 are empty; none is incomplete. The absence of incomplete cases follows from these two fully observed tables and must not be interpreted as an external missing-data evaluation. The correctness evidence for UNKNOWN handling comes from the explicit contract and separate boundary and state tests.

Supplement S2 retains the original same-origin construction table. Across all 1,280 sources, the median of the per-source construction medians is 38.550 microseconds for Atomic, 48.875 for SLA and 70.100 for Cover. The paired Cover/SLA geometric construction ratio is 1.45093, while SLA's sum of source medians is 31.18\% smaller. These are different summaries of the same paired observations; neither is a whole-session speedup. SLA is slower than Cover at 20 of the 1,280 individual source points, so the evidence does not show a pointwise timing guarantee.

The lower construction cost comes with lower same-source coverage. Across the 39,680 possible future source/query pairs, Atomic contains 2,478, SLA contains 4,936 and Cover contains 13,347. Pointwise inclusion agrees with Eq. (7). The difference is pronounced for Concrete: mean next-31 coverage is 0.0594 for Atomic, 0.9594 for SLA and 10.0797 for Cover. In Airfoil the corresponding means are 3.8125, 6.7531 and 10.7750. Thus SLA broadens the atomic region, but does not preserve the reuse range achieved by Cover.

\begin{figure}[!htbp]
\centering
\includegraphics[width=\textwidth,height=.64\textheight,keepaspectratio]{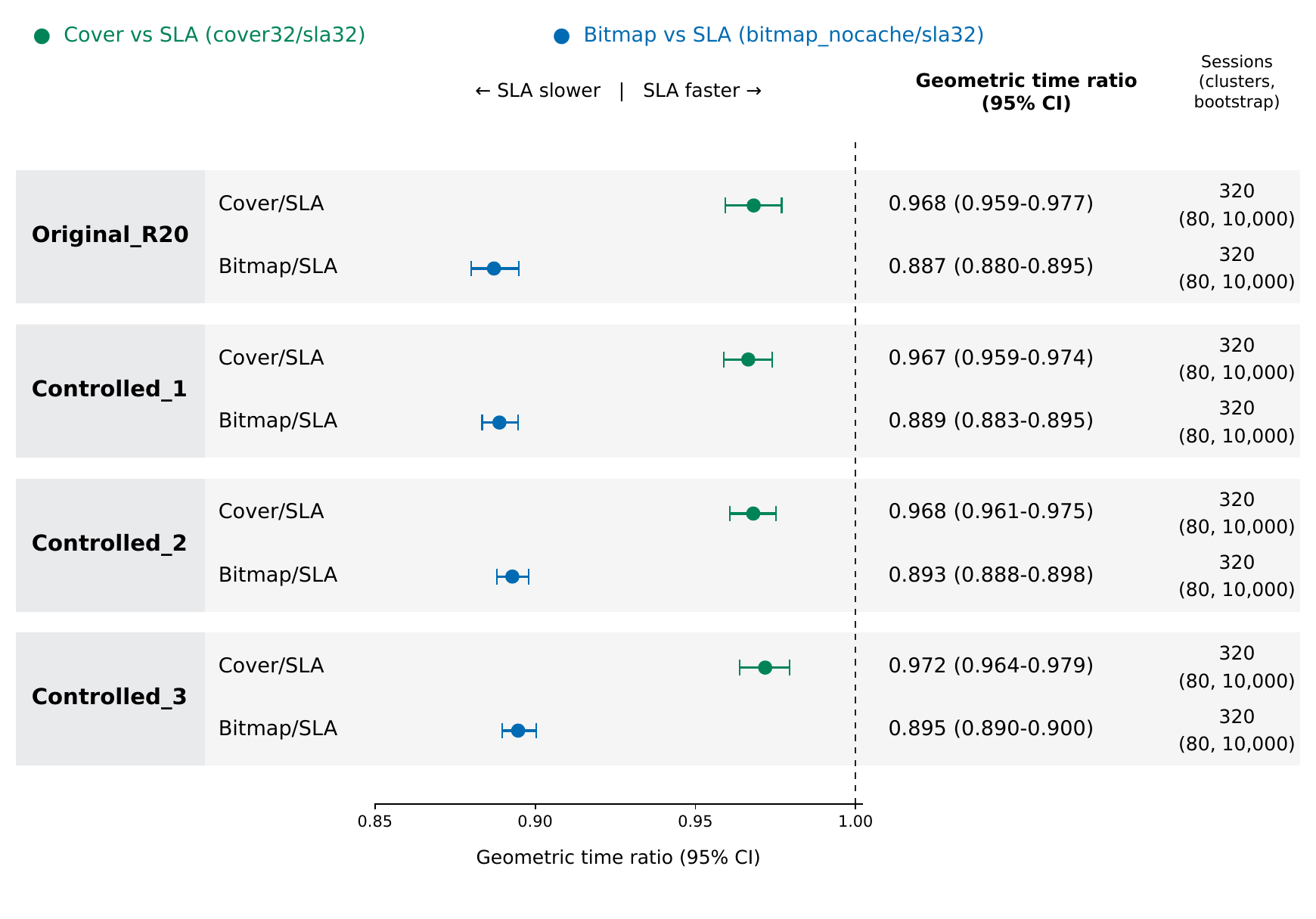}
\caption{Original and three separately launched full-matrix runs. Each interval is a 95\% conditional paired blocked bootstrap interval using 320 sessions, 80 replicate clusters and 10,000 draws with four workloads resampled together. They are not cross-run or industrial-population intervals. The dotted line denotes equal time.}
\label{fig:3}
\end{figure}

The original Cover/SLA geometric ratio remains 0.9682 [0.9593,0.9769]; SLA uses 3.10\% more summed session-median time than Cover and 12.94\% more than Bitmap. No aggregate improvement over Atomic was established either. The three additional runs retain the same adverse ordering (Table~\ref{tab:main-4}; Figure~\ref{fig:3}), even though summed same-origin construction medians fall by 40.15--40.22\% relative to Cover in their changed environment. The original 31.18\% construction saving is not overwritten by this newer range. Conditional intervals in different runs are shown separately; their endpoints are not treated as a cross-run confidence interval.

\begingroup
\small
\setlength{\tabcolsep}{4pt}
\begin{longtable}{@{}>{\raggedright\arraybackslash}p{0.094891\dimexpr\textwidth-10\tabcolsep\relax}>{\raggedright\arraybackslash}p{0.394161\dimexpr\textwidth-10\tabcolsep\relax}>{\raggedright\arraybackslash}p{0.167883\dimexpr\textwidth-10\tabcolsep\relax}>{\raggedright\arraybackslash}p{0.167883\dimexpr\textwidth-10\tabcolsep\relax}>{\raggedright\arraybackslash}p{0.175182\dimexpr\textwidth-10\tabcolsep\relax}@{}}
\caption{Original primary outcome and all controlled repeats; ratios above one favor SLA. Positive summed changes are slower.}\label{tab:main-4}\\
\toprule
\textbf{Run} & \textbf{Cover/SLA GM [95\% CI]} & \textbf{SLA sum vs Cover (\%)} & \textbf{SLA sum vs Bitmap (\%)} & \textbf{Construction saving (\%)} \\
\midrule
\endfirsthead
\multicolumn{5}{l}{\small\itshape Table \thetable{} (continued)}\\
\toprule
\textbf{Run} & \textbf{Cover/SLA GM [95\% CI]} & \textbf{SLA sum vs Cover (\%)} & \textbf{SLA sum vs Bitmap (\%)} & \textbf{Construction saving (\%)} \\
\midrule
\endhead
\midrule
\multicolumn{5}{r}{\small\itshape Continued on next page}\\
\endfoot
\bottomrule
\endlastfoot
Original & 0.9682 [0.9593, 0.9769] & +3.10 & +12.94 & 31.18 \\
1 & 0.9665 [0.9588, 0.9739] & +3.45 & +12.70 & 40.17 \\
2 & 0.9681 [0.9607, 0.9751] & +3.22 & +12.10 & 40.15 \\
3 & 0.9718 [0.9638, 0.9794] & +2.86 & +11.89 & 40.22 \\
\end{longtable}
\endgroup

\subsection*{Scale, permissions, locality and measured costs}
\label{sec:5-2}

\begin{figure}[!htbp]
\centering
\includegraphics[width=\textwidth,height=.64\textheight,keepaspectratio]{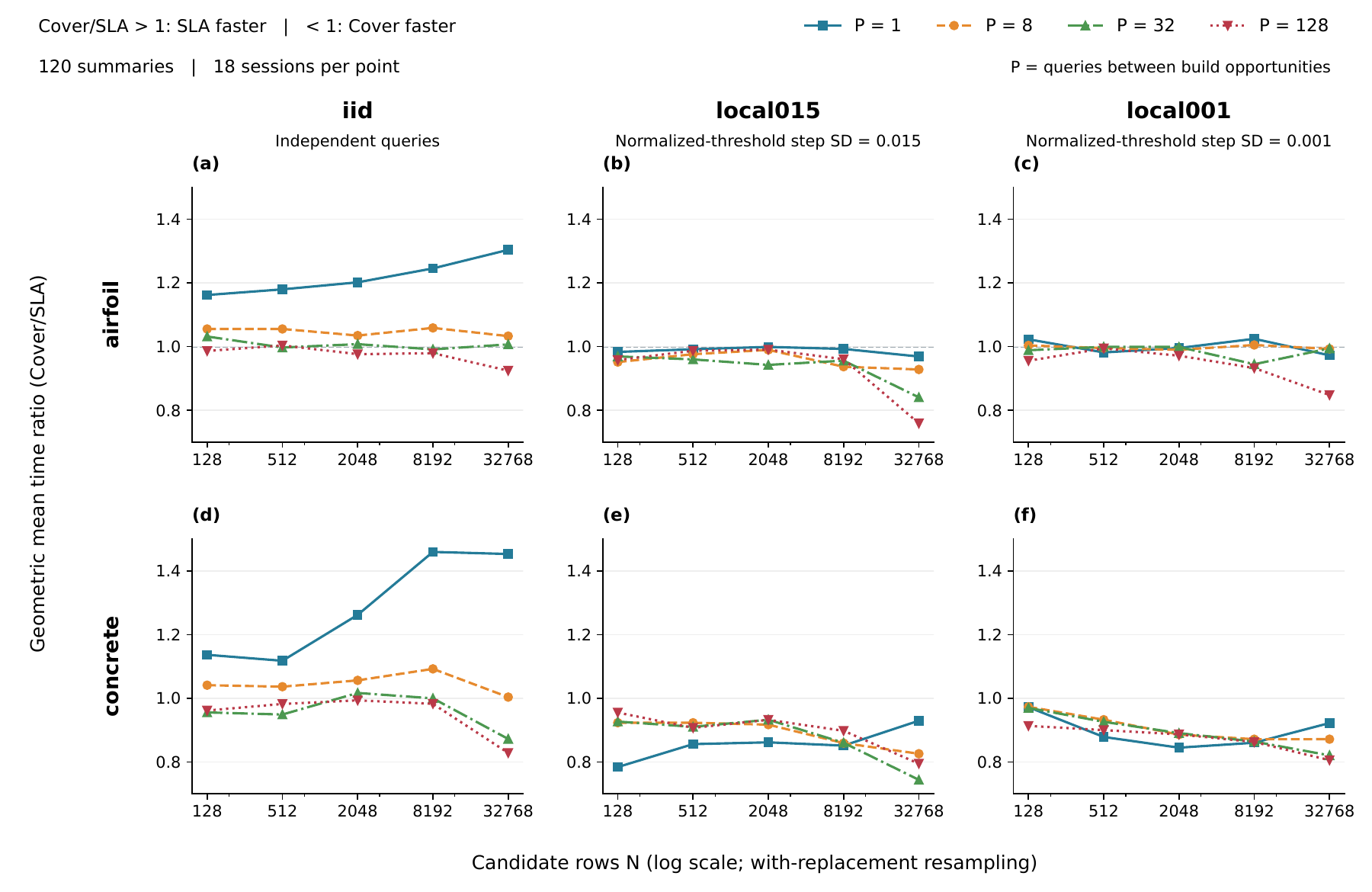}
\caption{Complete scale-period exploration of Cover/SLA, separated by dataset and query locality. Each displayed point is the geometric mean of 18 sessions: three k values \ensuremath{\times} two strata \ensuremath{\times} three seeds. Curves distinguish construction periods; candidate count is displayed on a logarithmic horizontal axis. The complete 720-configuration export also includes Bitmap contrasts; this aggregation does not establish significance or generalization to new physical records.}
\label{fig:4}
\end{figure}

Across all 720 exploratory configurations, the geometric ratio is above one in 295 cases for Cover/SLA, 271 for Bitmap/SLA and 372 for Bitmap/Cover. These are descriptive counts over three seeds per configuration, not significance-tested wins, and the first two counts are not a count of simultaneous superiority. Figure~\ref{fig:4} includes every size/period/locality cell; the machine-readable supplement preserves all 720 k- and stratum-resolved configurations. Table~\ref{tab:main-5} reports all marginal levels of locality, period and k, averaged across the remaining factors without selecting favourable cells.

\begingroup
\small
\setlength{\tabcolsep}{4pt}
\begin{longtable}{@{}>{\raggedright\arraybackslash}p{0.218978\dimexpr\textwidth-10\tabcolsep\relax}>{\raggedright\arraybackslash}p{0.145985\dimexpr\textwidth-10\tabcolsep\relax}>{\raggedright\arraybackslash}p{0.145985\dimexpr\textwidth-10\tabcolsep\relax}>{\raggedright\arraybackslash}p{0.240876\dimexpr\textwidth-10\tabcolsep\relax}>{\raggedright\arraybackslash}p{0.248175\dimexpr\textwidth-10\tabcolsep\relax}@{}}
\caption{Descriptive marginal geometric ratios and arithmetic mean counts per 128-query session. Each locality/k level pools 720 sessions; each period pools 540. No multiplicity-adjusted inference is made.}\label{tab:main-5}\\
\toprule
\textbf{Factor / level} & \textbf{Cover/SLA} & \textbf{Bitmap/SLA} & \textbf{Hits SLA / Cover} & \textbf{Builds SLA / Cover} \\
\midrule
\endfirsthead
\multicolumn{5}{l}{\small\itshape Table \thetable{} (continued)}\\
\toprule
\textbf{Factor / level} & \textbf{Cover/SLA} & \textbf{Bitmap/SLA} & \textbf{Hits SLA / Cover} & \textbf{Builds SLA / Cover} \\
\midrule
\endhead
\midrule
\multicolumn{5}{r}{\small\itshape Continued on next page}\\
\endfoot
\bottomrule
\endlastfoot
workload=iid & 1.0533 & 0.7820 & 6.31 / 17.50 & 35.22 / 32.45 \\
workload=local001 & 0.9347 & 1.2513 & 97.73 / 117.47 & 5.49 / 2.22 \\
workload=local015 & 0.9134 & 0.9498 & 45.37 / 78.85 & 18.90 / 9.51 \\
period=1 & 1.0271 & 0.8263 & 64.29 / 81.95 & 63.71 / 46.05 \\
period=8 & 0.9716 & 1.0201 & 55.37 / 77.05 & 11.39 / 8.93 \\
period=32 & 0.9399 & 1.0418 & 45.05 / 68.09 & 3.39 / 2.92 \\
period=128 & 0.9253 & 1.0328 & 34.50 / 58.01 & 1.00 / 1.00 \\
k=1 & 0.9920 & 0.9105 & 42.16 / 65.08 & 21.69 / 15.91 \\
k=5 & 0.9546 & 0.9650 & 49.45 / 71.60 & 20.04 / 14.75 \\
k=20 & 0.9496 & 1.0577 & 57.80 / 77.14 & 17.89 / 13.51 \\
\end{longtable}
\endgroup

Independent query coordinates make construction economy relatively more visible: iid has Cover/SLA = 1.0533 but Bitmap/SLA = 0.7820. Very local coordinates reverse the Bitmap contrast (local001: 1.2513), while Cover/SLA remains 0.9347. At P = 1, mean builds are 63.71 for SLA and 46.05 for Cover; at P = 128 both build once, but their hit counts still differ. Thus reducing build frequency does not remove the coverage trade-off. At N = 32,768, the marginal Bitmap/SLA ratio is 1.2197 whereas Cover/SLA is 0.9238. None of these margins identifies one certificate as best across all query families.

\begin{figure}[!htbp]
\centering
\includegraphics[width=\textwidth,height=.64\textheight,keepaspectratio]{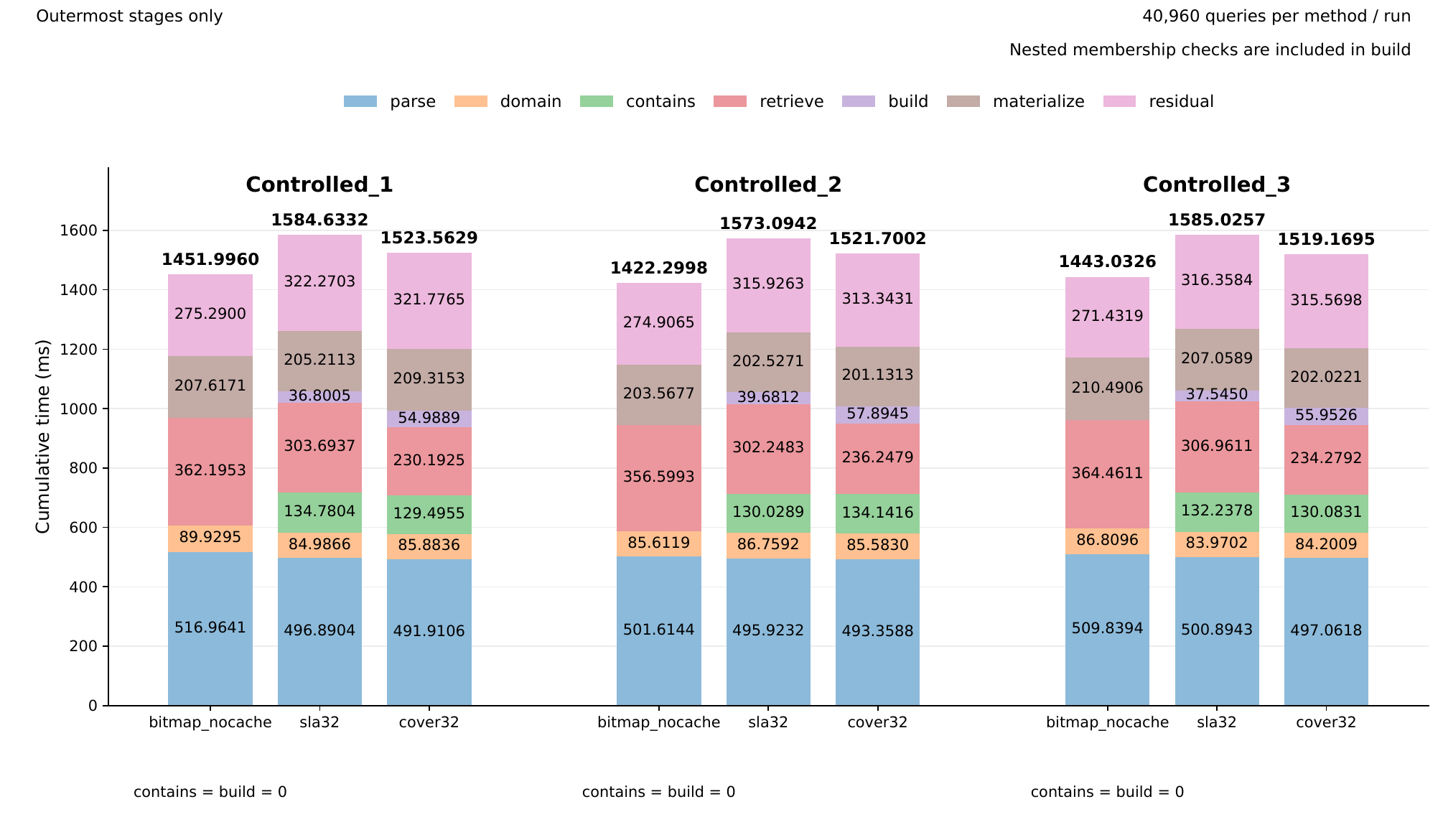}
\caption{Directly instrumented additive costs for all 320 original-matrix sessions in each repeat (40,960 queries per method per run). Six nonoverlapping stages plus residual sum to the same instrumented full-call total; residual includes timer/dispatch overhead. Nested construction checks remain in build. These perturbed measurements are not the primary uninstrumented timings and are not differences between unrelated medians.}
\label{fig:5}
\end{figure}

The direct first-repeat diagnostics give 36.80 ms of SLA construction against 54.99 ms for Cover, but 303.69 ms of SLA retrieval against 230.19 ms for Cover across 40,960 positions. Fresh report creation costs 205.21 and 209.32 ms, respectively. The corresponding full instrumented totals are 1,584.63 and 1,523.56 ms. All three repeats have 5,294 SLA hits versus 13,917 Cover hits, and 1,268 versus 1,203 builds. Figure~\ref{fig:5} shows the full three-run accounting rather than attributing the complete performance gap to constructor medians alone.

\begin{figure}[!htbp]
\centering
\includegraphics[width=\textwidth,height=.64\textheight,keepaspectratio]{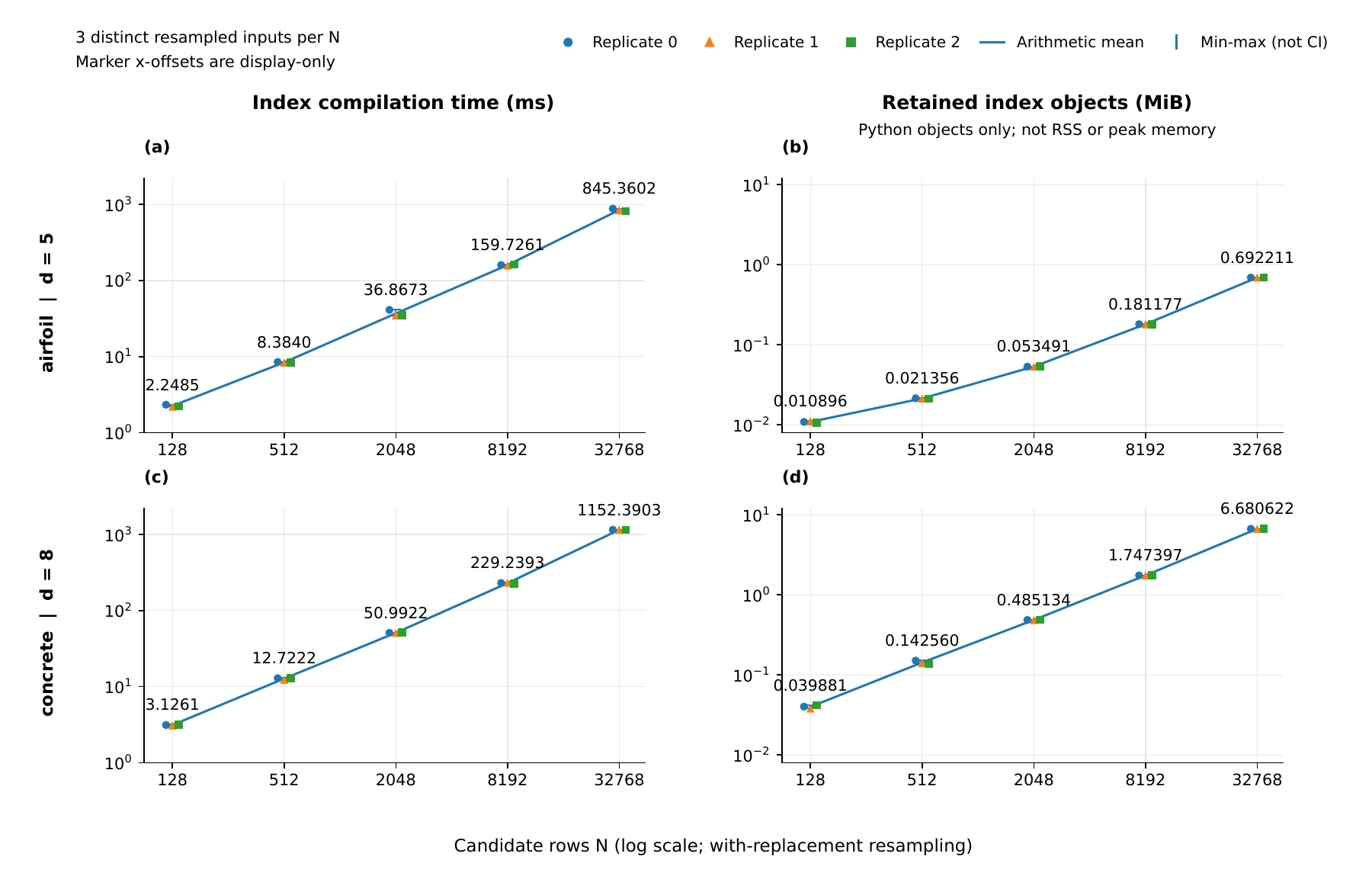}
\caption{All 30 shared-catalogue preparation observations, three seeds per dataset and size. Left: compile time. Right: retained Python column-index object bytes, excluding feature arrays and whole-process RSS. Display offsets distinguish seeds, not sizes. These are not OS-cold-cache or per-method exclusive costs.}
\label{fig:6}
\end{figure}

At N = 32,768, Airfoil compilation takes 818.34--880.24 ms and its index retains 0.69210--0.69244 MiB; Concrete takes 1,148.94--1,155.93 ms with 6.68018--6.68146 MiB (Figure~\ref{fig:6}). Separate feature-array sizes are 1.28125 and 2.03125 MiB. Repeated values and bitmap representation explain why retained memory does not follow N alone. The API comparison excludes this shared preparation; an application with short-lived catalogues must account for it separately.

\subsection*{Mechanics and state-stratified functional results}
\label{sec:5-3}

The unchanged exact-traction reference achieves finest-grid hole-stress error 0.00182\% and profile NRMSE 0.01396\%. Added displacement relative L2 errors decrease from 2.6044\ensuremath{\times}10\textsuperscript{-4} to 1.9570\ensuremath{\times}10\textsuperscript{-5} to 1.4139\ensuremath{\times}10\textsuperscript{-6} globally, and from 7.4706\ensuremath{\times}10\textsuperscript{-4} to 5.4949\ensuremath{\times}10\textsuperscript{-5} to 3.9472\ensuremath{\times}10\textsuperscript{-6} near the hole. The finest values correspond to 0.0001414\% and 0.0003947\%, well below the 1\% criterion. Independent constitutive differentiation reproduces analytical stresses to relative errors between 7.0\ensuremath{\times}10\textsuperscript{-11} and 2.2\ensuremath{\times}10\textsuperscript{-7} across four finite-difference steps. This checks the correct plane-stress reference, not the different finite-plate uniform-loading solution.

\begingroup
\small
\setlength{\tabcolsep}{4pt}
\begin{longtable}{@{}>{\raggedright\arraybackslash}p{0.062500\dimexpr\textwidth-14\tabcolsep\relax}>{\raggedright\arraybackslash}p{0.098214\dimexpr\textwidth-14\tabcolsep\relax}>{\raggedright\arraybackslash}p{0.098214\dimexpr\textwidth-14\tabcolsep\relax}>{\raggedright\arraybackslash}p{0.205357\dimexpr\textwidth-14\tabcolsep\relax}>{\raggedright\arraybackslash}p{0.178571\dimexpr\textwidth-14\tabcolsep\relax}>{\raggedright\arraybackslash}p{0.178571\dimexpr\textwidth-14\tabcolsep\relax}>{\raggedright\arraybackslash}p{0.178571\dimexpr\textwidth-14\tabcolsep\relax}@{}}
\caption{All nine fine-mesh finite-plate candidates; stress change is medium-to-fine relative change.}\label{tab:main-6}\\
\toprule
\textbf{ID} & \textbf{W (mm)} & \textbf{t (mm)} & \textbf{Volume (mm\textsuperscript{3})} & \textbf{Hole \ensuremath{\sigma_{xx}} (MPa)} & \textbf{\ensuremath{\bar{u}} (mm)} & \textbf{Stress change (\%)} \\
\midrule
\endfirsthead
\multicolumn{7}{l}{\small\itshape Table \thetable{} (continued)}\\
\toprule
\textbf{ID} & \textbf{W (mm)} & \textbf{t (mm)} & \textbf{Volume (mm\textsuperscript{3})} & \textbf{Hole \ensuremath{\sigma_{xx}} (MPa)} & \textbf{\ensuremath{\bar{u}} (mm)} & \textbf{Stress change (\%)} \\
\midrule
\endhead
\midrule
\multicolumn{7}{r}{\small\itshape Continued on next page}\\
\endfoot
\bottomrule
\endlastfoot
1 & 80 & 4 & 62743.36 & 203.016 & 0.031651 & 0.003072 \\
2 & 80 & 6 & 94115.04 & 135.344 & 0.021101 & 0.003072 \\
3 & 80 & 8 & 125486.73 & 101.508 & 0.015826 & 0.003072 \\
4 & 100 & 4 & 78743.36 & 157.796 & 0.024988 & 0.005118 \\
5 & 100 & 6 & 118115.04 & 105.197 & 0.016658 & 0.005118 \\
6 & 100 & 8 & 157486.73 & 78.898 & 0.012494 & 0.005118 \\
7 & 120 & 4 & 94743.36 & 129.816 & 0.020650 & 0.008018 \\
8 & 120 & 6 & 142115.04 & 86.544 & 0.013767 & 0.008018 \\
9 & 120 & 8 & 189486.73 & 64.908 & 0.010325 & 0.008018 \\
\end{longtable}
\endgroup

The largest normalized force, moment and energy residuals over positive-load runs are 2.56448 \ensuremath{\times} 10\textsuperscript{-13}, 7.84377 \ensuremath{\times} 10\textsuperscript{-14} and 2.09757 \ensuremath{\times} 10\textsuperscript{-13}, respectively. Native-versus-reconstructed integration-point stress NRMSE is at most 5.48103 \ensuremath{\times} 10\textsuperscript{-14}. Zero-load absolute checks pass. For the stored half-load solution, displacement and stress are one half, and energy one quarter, of the corresponding full-load fields with zero relative deviation in the stored floating-point comparison. These are strong consistency and linearity checks within the declared model, not evidence that a physical connection behaves linearly.

\begin{figure}[!htbp]
\centering
\includegraphics[width=\textwidth,height=.64\textheight,keepaspectratio]{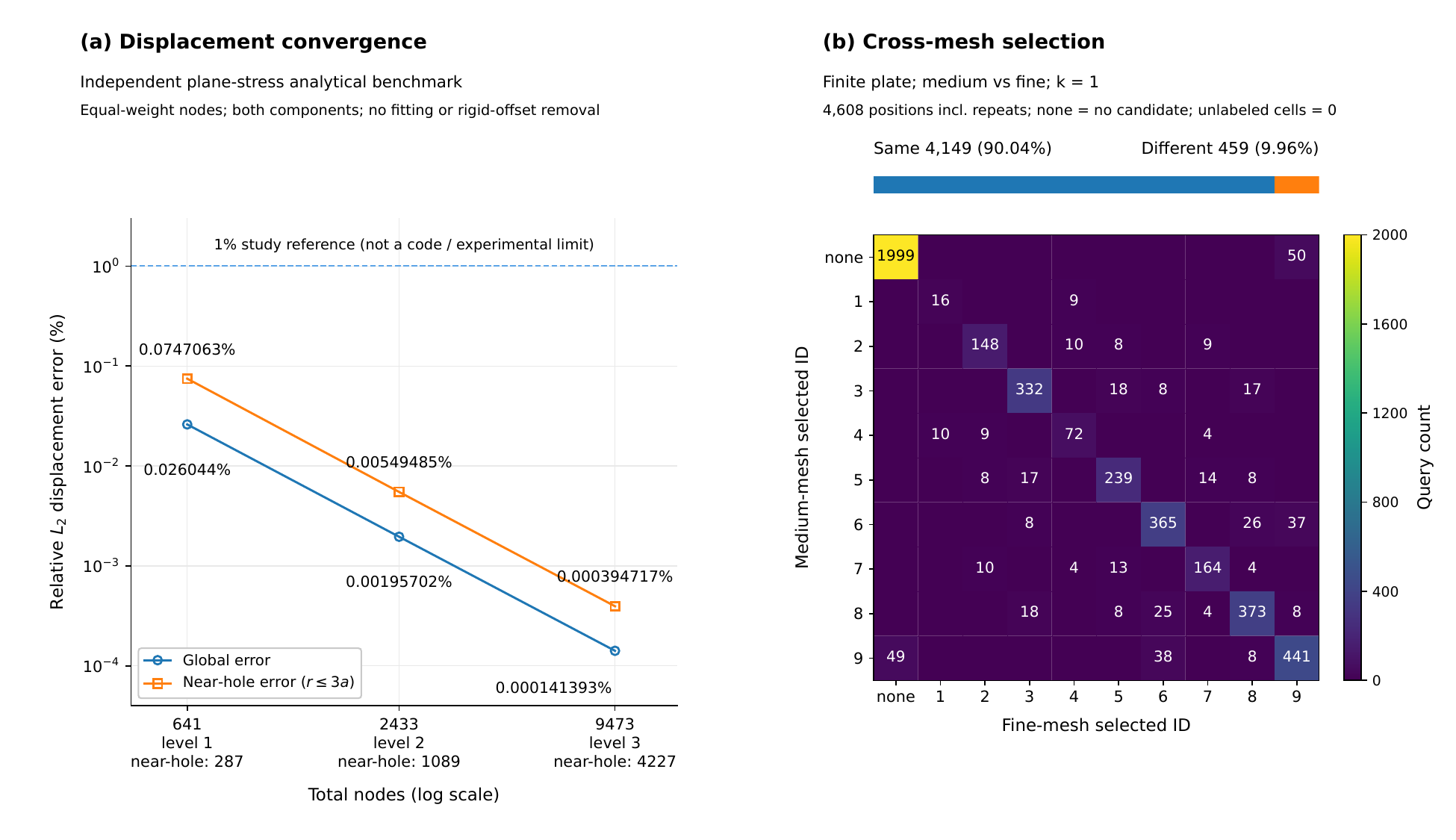}
\caption{Independent plane-stress displacement check and mesh-dependent selection. Left: unweighted nodal vector relative L2 error, displayed as percent, globally and at r \ensuremath{\leq} 3a over three meshes. Right: counts of actual medium/fine selected IDs over all 4,608 positions, including empty (labelled none, encoded 0); 459 lie off the diagonal. These are actual mesh comparisons, not the 1,323 coordinate-envelope ambiguities.}
\label{fig:7}
\end{figure}

The original 49-query grid exercises only four feasible sets and selected IDs \{1,2,4,8\}, with zero envelope ambiguities; its 245 reports and the 270 original boundary reports still agree. In contrast, the added 4,608 positions exercise 17 feasible sets and all nine selected IDs, including 2,048 empty results (Table~\ref{tab:main-7}). Geometry constraints change 2,832 selections relative to inactive geometry. Medium/fine selected IDs actually differ at 459 positions. Optimistic/pessimistic coordinate envelopes are ambiguous at 1,323 positions (27 near-attribute and 1,296 crossed-geometry positions), a different, more conservative count. Figure~\ref{fig:7} shows both displacement convergence and the actual selection transitions. All 23,040 new full-report comparisons agree with the scalar reference; that agreement concerns a chosen attribute table and cannot eliminate its mesh sensitivity.

A concrete query illustrates the added mechanics-to-selection link. With stress \ensuremath{\leq} 150 MPa and \ensuremath{\bar{u}} \ensuremath{\leq} 0.025 mm, candidate 2 (W = 80 mm, t = 6 mm) is selected: volume 94,115.04 mm\textsuperscript{3}, hole stress 135.344 MPa and \ensuremath{\bar{u}} = 0.021101 mm. The lower-volume candidate 1 fails both limits (203.016 MPa; 0.031651 mm). Candidate 4 also precedes candidate 2 by volume and passes the displacement limit (0.024988 mm), but its stress of 157.796 MPa excludes it. The returned candidate is therefore minimal in the declared finite ordering, not a continuous global design optimum or a code-approved connector.

The original bolt workload is retained exactly: 66 answered, 4,306 empty and 1,088 incomplete positions, with 27,300 matching method/reference reports. The additional attainable-capacity workload yields 994 answered, 300 empty and 185 incomplete positions, spanning 50 distinct ordered selected sets. Its 7,395 comparisons all agree, including 925 incomplete-state comparisons, with quarantined IDs 12 and 23 unresolved where rank-relevant. Thus the new exercise adds positive and near-boundary retrieval coverage without discarding the earlier mostly empty workload or concealing source uncertainty. It measures execution of declared finite-record predicates, not design resistance.

\begingroup
\small
\setlength{\tabcolsep}{4pt}
\begin{longtable}{@{}>{\raggedright\arraybackslash}p{0.204380\dimexpr\textwidth-12\tabcolsep\relax}>{\raggedright\arraybackslash}p{0.145985\dimexpr\textwidth-12\tabcolsep\relax}>{\raggedright\arraybackslash}p{0.138686\dimexpr\textwidth-12\tabcolsep\relax}>{\raggedright\arraybackslash}p{0.138686\dimexpr\textwidth-12\tabcolsep\relax}>{\raggedright\arraybackslash}p{0.153285\dimexpr\textwidth-12\tabcolsep\relax}>{\raggedright\arraybackslash}p{0.218978\dimexpr\textwidth-12\tabcolsep\relax}@{}}
\caption{Complete state accounting; every comparison agrees. Positions include repeated requests and are not independent physical tests.}\label{tab:main-7}\\
\toprule
\textbf{Case} & \textbf{Positions} & \textbf{Answered} & \textbf{Empty} & \textbf{Incomplete} & \textbf{Report comparisons} \\
\midrule
\endfirsthead
\multicolumn{6}{l}{\small\itshape Table \thetable{} (continued)}\\
\toprule
\textbf{Case} & \textbf{Positions} & \textbf{Answered} & \textbf{Empty} & \textbf{Incomplete} & \textbf{Report comparisons} \\
\midrule
\endhead
\midrule
\multicolumn{6}{r}{\small\itshape Continued on next page}\\
\endfoot
\bottomrule
\endlastfoot
NIST original & 5,460 & 66 & 4,306 & 1,088 & 27,300 \\
NIST supplement & 1,479 & 994 & 300 & 185 & 7,395 \\
FE original grid & 49 & 49 & 0 & 0 & 245 \\
FE original probes & 54 & 52 & 2 & 0 & 270 \\
FE supplement & 4,608 & 2,560 & 2,048 & 0 & 23,040 \\
\end{longtable}
\endgroup

\subsection*{Reproducibility checks}
\label{sec:5-4}

The accompanying versioned computation archive includes all original per-call timings, all 3,120 supplemental session arrays, diagnostic counters/stages, frozen inputs and query seeds, source manifests, environment lock and recomputation scripts. After a fresh isolated Python 3.12 environment was installed from the lock file, all original and new statistical summaries were recomputed, including blocked intervals; scientific fields matched. Core correctness tests, the counterexample, Q8 self-tests, FE/NIST boundary reports and all five new figures also rebuilt. Two representative FE solves---medium exact-traction verification and the fine central finite plate---were rerun, and all native numeric fields matched the retained fields in that environment. An independent execution from the actual extracted ZIP repeated these checks. This is successful same-host clean-environment recomputation, not fresh performance measurement, full re-solving of all 23 FE cases or cross-platform verification. Supplement S4 provides the command sequence and provenance.

\section*{Discussion and limitations}
\label{sec:6}

Construction economy, same-source containment and online speed are different claims. The first is supported for SLA against Cover; the primary full-API improvement is not. The controlled repeats preserve that distinction, while direct instrumentation shows lower construction cost alongside greater retrieval work and unavoidable current-report costs. The deterministic counterexample additionally shows why source-level containment cannot order every retained online state under a nonaccumulating permission policy. A service should therefore measure the complete response contract and compare a strong uncached backend before optimizing certificate geometry alone.

The sensitivity matrix extends the evidence across size, permissions, requested count and locality, but it remains exploratory. Its resampled empirical support, shared seed prefixes, catalogue-derived positive anchors and synthetic threshold motion do not model unseen engineering projects or measurement noise. Marginal gains against one comparator coexist with losses against another. Adaptive selection of a retrieval/certificate policy is a plausible future direction, but no learned or adaptive selector is implemented or validated here. A new held-out workload would be required to test any policy chosen from this matrix.

The engineering additions strengthen two narrow links. Analytical displacement verification checks the correct plane-stress boundary problem independently of the solver, and adversarially close thresholds expose decision instability hidden by the coarse original grid. State stratification increases bolt-record positive coverage while retaining UNKNOWN quarantine. Neither exact report agreement nor a small deterministic mesh error validates uncertain material properties, experimental noise or an actual joint. Nine parameter variants remain one model family. The NIST window minima are observations under a finite archive rule, not statistically justified design resistances.

All fields in the two timed tables are known. The bolt-record exercise adds real source-conflict quarantine, but not a realistic probability distribution of missingness or measurement error. Exact report agreement does not establish robustness to uncertain forces, uncertain displacements or adversarial noise. The model assumes fixed rank and monotone per-coordinate inequalities. Query-dependent rankings, coupled nonlinear predicates, tolerance stacks, probabilistic constraints and incremental index maintenance require new algorithms or a separately justified reduction. Resetting on epoch change is conservative invalidation, not an efficient dynamic-update contribution.

The literature comparison establishes where the present problem overlaps mature approaches, but is not exhaustive novelty clearance. In particular, Ehlers's doctoral thesis Top-k Semantic Caching \cite{ref6} was identified through institutional metadata, while its full-method comparison remains incomplete. We do not infer that its unread methods lack the endpoint rule studied here. Nor have we benchmarked the complete cited range-query and caching systems or modern external database implementations. The restricted lower-bound characterization and controlled observations should be evaluated at that level of theoretical and empirical strength.

The elastic component has no bearing contact, friction, pretension, plasticity, fatigue or failure model. We do not claim an independent nonlinear joint test, exact B-rep assembly/collision check, comprehensive code compliance or human structural-engineer blind review. The prior screw-record source with unresolved identity/force conflicts remains excluded rather than pooled into engineering evidence. Fresh-environment checks establish reproducibility of the supplied computations on this host; background load, changed libraries/warmup and a single CPU architecture still limit performance transportability. Genuine author contributions and approval remain necessary before submission.

\section*{Conclusions}
\label{sec:7}

SLA provides exact fixed-rank identity reuse under the stated completeness and digital-domain contract, with a restricted fixed-upper-endpoint characterization. The original 31.18\% construction saving does not yield primary full-API acceleration: Cover/SLA = 0.9682 [0.9593,0.9769], and Bitmap is faster still. Three separately launched complete repeats preserve the adverse ordering. The full 720-configuration exploration and additive diagnostics show conditional trade-offs rather than one universally superior method. For transparency, candidate construction was explored and refined through ZiYor, while the named authors specified, implemented and evaluated the method.

The mechanics-to-query chain now includes an independent plane-stress displacement reference, all nine finite-catalogue selections, empty results and active geometry constraints. Its 459 actual medium/fine selection differences demonstrate why coarse-grid agreement cannot establish robust selection. The supplementary bolt exercise supplies 994 answered and 185 explicitly incomplete positions without concealing original negative coverage. A tested raw-data computation package supports audit and reuse; physical connection qualification, unrestricted novelty and universal selection speedup remain outside the demonstrated result.

\section*{Declarations}

Funding. This research received no external funding.

Competing interests. The authors declare no competing interests.

Data availability. The original Airfoil Self-Noise and Concrete Compressive Strength data are available from the UCI repository \cite{ref10,ref11}, under CC BY 4.0. Original attributions and all duplicate rows are retained. The original bolt workbooks are available through NIST DOI 10.18434/M33T0T \cite{ref16}. The supplemental source manifest records original byte hashes and the derived, source-conflicted record treatment; our derived predicates are not endorsed NIST design values. Raw third-party papers are not redistributed. The added elastic benchmark is attributed to public documentation \cite{ref18,ref19}, with exact outer tractions explicitly distinguished from the documented finite-boundary setup. Author-generated model inputs, native FE arrays and derived catalogue records accompany the local case package.

Code and computational records. The local reviewer archive supplies original and supplementary raw timings, unchanged screening core, independent references, Q8 solver, locked dependencies, source hashes and reproducible statistics/figure/engineering entry points. Fresh-environment recomputation and two representative FE re-solves have actually passed, with the limits described in Section~\nameref{sec:5-4}. No public repository, archive DOI, cross-platform result or approved open-source licence is asserted. Third-party papers are not redistributed; public release requires rights-holder approval.

Author contributions. Individual CRediT roles remain pending genuine author confirmation; this preparation draft does not assign unverified roles to the named authors.

\section*{Acknowledgements and AI disclosure}

OpenAI Codex assisted research design, source investigation, software, experiment execution and checking, analysis, visualization and writing. Its use was substantive, not limited to language editing. The exact underlying model version is unavailable in the retained records. AI-assisted computational review is not human scientific peer review. The named authors must personally verify the scientific content, source rights and genuine contributions and approve submission; this preparation draft does not certify that those human actions have occurred.

The authors report that the selected-lower, atomic-upper (SLA) certificate construction was explored and refined through the self-evolving AI system ZiYor. The named authors then specified, implemented and evaluated the resulting method, including the reported experiments, mechanics checks and reproducibility package. ZiYor is not an author and did not independently validate the data, code or scientific conclusions. All scientific content and the final manuscript remain the responsibility of the named authors.

\clearpage
% Verified mechanical-screening bibliography; stable manuscript keys ref1--ref21.
% Updated 2026-09-27. Full-method inspection of ref6 remains unresolved in the manuscript.

\clearpage
\section*{Supplementary information}
\setcounter{table}{0}
\renewcommand{\thetable}{S\arabic{table}}
\renewcommand{\theHtable}{supplement.\arabic{table}}
% Auto-generated from canonical JSON; see conversion_audit.json.

\section*{S1. Six-axis closest-method comparison}

The numbered references are the same as in the main manuscript. Tables~\ref{tab:supplement-1} and~\ref{tab:supplement-2} form one six-axis comparison, split only for legibility. A feature not present in the inspected formulation is not claimed absent from every extension. No external system was implemented and timed for this paper. The distinction from previous methods is a bounded fixed-rank/report-contract study, not priority for semantic caching, safe regions, negative-result reuse or generic cost accounting.

\begingroup
\small
\setlength{\tabcolsep}{4pt}
\begin{longtable}{@{}>{\raggedright\arraybackslash}p{0.145985\dimexpr\textwidth-8\tabcolsep\relax}>{\raggedright\arraybackslash}p{0.262774\dimexpr\textwidth-8\tabcolsep\relax}>{\raggedright\arraybackslash}p{0.248175\dimexpr\textwidth-8\tabcolsep\relax}>{\raggedright\arraybackslash}p{0.343066\dimexpr\textwidth-8\tabcolsep\relax}@{}}
\caption{Cached object, query changes and region construction.}\label{tab:supplement-1}\\
\toprule
\textbf{Work} & \textbf{Cached object} & \textbf{Allowed query change} & \textbf{Region / bound} \\
\midrule
\endfirsthead
\multicolumn{4}{l}{\small\itshape Table \thetable{} (continued)}\\
\toprule
\textbf{Work} & \textbf{Cached object} & \textbf{Allowed query change} & \textbf{Region / bound} \\
\midrule
\endhead
\midrule
\multicolumn{4}{r}{\small\itshape Continued on next page}\\
\endfoot
\bottomrule
\endlastfoot
Huang et al. \cite{ref2} & Spatial top-k answer, safe region and traversal information & Location; keywords and k fixed & Dominance-region intersection, indexed pruning and conservative polygons \\
McClain et al. \cite{ref3} & Bit-vectors for query segments & Range/point segments combined by AND/OR & Range-restricted MaximizeCoverage selects cached vectors and remainder \\
Xie et al. \cite{ref5} & Top-k views with attributes, ranks and scores & Scoring function and k & LP unseen-score threshold; LPTA+ basis reuse; IV-index \\
Keller--Basu \cite{ref7} & Relational tuples and predicate descriptions & Selection, projection and join with required keys & Conservative client descriptions; liberal server descriptions \\
Dar et al. \cite{ref8} & Disjoint semantic regions, tuples and constraint formulae & Selection predicates on a relation & Probe/remainder splitting; repartition and coalescence \\
Ehlers \cite{ref6} & Not method-verified & Not method-verified & Full thesis unavailable \\
Present model & One ordered identity list and threshold certificate & Upper bounds; catalogue, rank, k and predicate family fixed & Selected maxima and atomic successors; competitor witnesses for Cover \\
\end{longtable}
\endgroup

\begingroup
\small
\setlength{\tabcolsep}{4pt}
\begin{longtable}{@{}>{\raggedright\arraybackslash}p{0.145985\dimexpr\textwidth-8\tabcolsep\relax}>{\raggedright\arraybackslash}p{0.277372\dimexpr\textwidth-8\tabcolsep\relax}>{\raggedright\arraybackslash}p{0.262774\dimexpr\textwidth-8\tabcolsep\relax}>{\raggedright\arraybackslash}p{0.313869\dimexpr\textwidth-8\tabcolsep\relax}@{}}
\caption{Missing information, current result processing and maintenance.}\label{tab:supplement-2}\\
\toprule
\textbf{Work} & \textbf{Missing information} & \textbf{Current result processing} & \textbf{Maintenance} \\
\midrule
\endfirsthead
\multicolumn{4}{l}{\small\itshape Table \thetable{} (continued)}\\
\toprule
\textbf{Work} & \textbf{Missing information} & \textbf{Current result processing} & \textbf{Maintenance} \\
\midrule
\endhead
\midrule
\multicolumn{4}{r}{\small\itshape Continued on next page}\\
\endfoot
\bottomrule
\endlastfoot
Huang et al. \cite{ref2} & Located/text-described objects; not field-level UNKNOWN & Client membership; server recomputation outside region & Replace answer/region on exit; not the present token policy \\
McClain et al. \cite{ref3} & Uncovered vectors fetched; not field-level UNKNOWN & Merge cached/remainder vectors; fetch tuples & Indiscriminate insertion; CLOCK capacity replacement \\
Xie et al. \cite{ref5} & Certain answers over score-consistent relations without base access & Score candidates under current query; retain provably certain tuples & Inspected sections address query evaluation/basis reuse, not this single-certificate policy \\
Keller--Basu \cite{ref7} & Cache completeness differs from database NULL/our missing features & Cache evaluation or server retrieval; query trimming & Update notifications; predicate maintenance; cost-based reclamation \\
Dar et al. \cite{ref8} & Uncached portions fetched; not this UNKNOWN contract & Combine cache and remainder tuples & LRU/distance region value; updates excluded from simulation \\
Ehlers \cite{ref6} & Not method-verified & Not method-verified & Not method-verified \\
Present model & Earlier unexcluded missing records block completeness/certification & Fresh margins and nested records on every call & Permission every P calls, no accumulation; eligible-miss replacement and epoch invalidation \\
\end{longtable}
\endgroup

Huang et al.: Sect. 2.1, Definitions 1--3; Sect. 3.2; Algorithm 2; Sects. 4.3--4.4 and 5.1 (printed pp. 933, 935, 937--938). McClain et al.: Sects. 3.1--3.4, Algorithm 1 and Sect. 4.3 (PDF pp. 3--5, 7). Xie et al.: Sect. 2, Definitions 1--2; Sect. 3.2, Algorithm 2; Sect. 3.3; Sect. 4.1 (printed pp. 491, 493--494). Keller and Basu: Sects. 3.1--3.2, Definitions 1--4; Sect. 4; Sects. 5.1--5.5 (printed pp. 38--39, 43--44). Dar et al.: Sects. 2.4, 3.1--3.4 and 4.2 (printed pp. 333--335). These locators support the paraphrases, not a declaration of exhaustive novelty.

The institutional listing identifies Ehlers (2015), Top-k Semantic Caching. The institutional full-text record was inaccessible behind an access challenge during this revision; an author-provided abstract is not a substitute for method chapters. No unavailable algorithm properties are inferred. The unresolved closest-thesis comparison remains an explicit residual gap. No third-party full-text papers are redistributed in the reviewer archive.

\section*{S2. Workload, measurement and complete outputs}

\begingroup
\small
\setlength{\tabcolsep}{4pt}
\begin{longtable}{@{}>{\raggedright\arraybackslash}p{0.136000\dimexpr\textwidth-8\tabcolsep\relax}>{\raggedright\arraybackslash}p{0.328000\dimexpr\textwidth-8\tabcolsep\relax}>{\raggedright\arraybackslash}p{0.136000\dimexpr\textwidth-8\tabcolsep\relax}>{\raggedright\arraybackslash}p{0.400000\dimexpr\textwidth-8\tabcolsep\relax}@{}}
\caption{Original native units and ranges; all source rows and source precision retained.}\label{tab:supplement-3}\\
\toprule
\textbf{Dataset} & \textbf{Feature} & \textbf{Unit} & \textbf{Observed range} \\
\midrule
\endfirsthead
\multicolumn{4}{l}{\small\itshape Table \thetable{} (continued)}\\
\toprule
\textbf{Dataset} & \textbf{Feature} & \textbf{Unit} & \textbf{Observed range} \\
\midrule
\endhead
\midrule
\multicolumn{4}{r}{\small\itshape Continued on next page}\\
\endfoot
\bottomrule
\endlastfoot
Airfoil & Frequency & Hz & 200--20,000 \\
Airfoil & Attack angle & degree & 0--22.2 \\
Airfoil & Chord length & m & 0.0254--0.3048 \\
Airfoil & Free-stream speed & m/s & 31.7--71.3 \\
Airfoil & Displacement thickness & m & 0.000400682--0.0584113 \\
Concrete & Cement & kg/m\textsuperscript{3} & 102--540 \\
Concrete & Blast-furnace slag & kg/m\textsuperscript{3} & 0--359.4 \\
Concrete & Fly ash & kg/m\textsuperscript{3} & 0--200.1 \\
Concrete & Water & kg/m\textsuperscript{3} & 121.75--247 \\
Concrete & Superplasticizer & kg/m\textsuperscript{3} & 0--32.2 \\
Concrete & Coarse aggregate & kg/m\textsuperscript{3} & 801--1,145 \\
Concrete & Fine aggregate & kg/m\textsuperscript{3} & 594--992.6 \\
Concrete & Age & day & 1--365 \\
\end{longtable}
\endgroup

Each dataset/k/stratum/replicate combination has a deterministic NumPy PCG64 stream. Its integer seed is obtained from the first eight SHA-256 digest bytes, interpreted little-endian, of the UTF-8 label R20-public-transfer-v1|dataset|k|stratum|replicate. Dataset strings are airfoil and concrete, strata are broad and positive, and replicate is a decimal integer from zero to nine. Independent anchor, method-order and constructor-order streams append |anchor, |order and |construct. These labels identify reproducible randomization, not selection based on measured outcomes.

Query generation consumes PCG64 draws in the fixed iid/local/shuffled/jumps order described in main Section~\nameref{sec:4-1}. Broad and positive mappings are main Eq. (9). The anchor uses original rows without replacement; paired orders preserve their shared source/anchor dependence. For the original aggregate bootstrap, use the label R20-public-transfer-v1|aggregate-bootstrap, the same eight-byte little-endian SHA-256 mapping and PCG64. Ten indices are sampled with replacement inside each of eight dataset/k/stratum blocks and jointly applied to the four orders; 10,000 draws and linear quantiles give each conditional interval. Group-specific labels replace the replicate field with group and append |bootstrap. Other contrasts/subgroups are exploratory without multiplicity correction.

\begingroup
\small
\setlength{\tabcolsep}{4pt}
\begin{longtable}{@{}>{\raggedright\arraybackslash}p{0.200000\dimexpr\textwidth-10\tabcolsep\relax}>{\raggedright\arraybackslash}p{0.200000\dimexpr\textwidth-10\tabcolsep\relax}>{\raggedright\arraybackslash}p{0.200000\dimexpr\textwidth-10\tabcolsep\relax}>{\raggedright\arraybackslash}p{0.200000\dimexpr\textwidth-10\tabcolsep\relax}>{\raggedright\arraybackslash}p{0.200000\dimexpr\textwidth-10\tabcolsep\relax}@{}}
\caption{Retained original same-source construction and coverage; median of source medians, not session speed.}\label{tab:supplement-4}\\
\toprule
\textbf{Scope} & \textbf{Method} & \textbf{Median (\ensuremath{\mu}s)} & \textbf{Mean next-31 coverage} & \textbf{Total coverage} \\
\midrule
\endfirsthead
\multicolumn{5}{l}{\small\itshape Table \thetable{} (continued)}\\
\toprule
\textbf{Scope} & \textbf{Method} & \textbf{Median (\ensuremath{\mu}s)} & \textbf{Mean next-31 coverage} & \textbf{Total coverage} \\
\midrule
\endhead
\midrule
\multicolumn{5}{r}{\small\itshape Continued on next page}\\
\endfoot
\bottomrule
\endlastfoot
Airfoil & Atomic & 36.500 & 3.8125 & 2,440/19,840 \\
Airfoil & SLA & 47.550 & 6.7531 & 4,322/19,840 \\
Airfoil & Cover & 64.025 & 10.7750 & 6,896/19,840 \\
Concrete & Atomic & 41.025 & 0.0594 & 38/19,840 \\
Concrete & SLA & 49.925 & 0.9594 & 614/19,840 \\
Concrete & Cover & 77.675 & 10.0797 & 6,451/19,840 \\
All & Atomic & 38.550 & 1.9359 & 2,478/39,680 \\
All & SLA & 48.875 & 3.8563 & 4,936/39,680 \\
All & Cover & 70.100 & 10.4273 & 13,347/39,680 \\
\end{longtable}
\endgroup

Original same-origin construction contains 1,280 locations, three constructors and ten timings per constructor. Source coverage is membership of the next 31 fixed queries; it does not reuse online states. The three new original-matrix processes repeat the sealed 320-session queries and original scalar reports. Each has 2,048,000 main calls plus 204,800 diagnostic positions. Warmup uses a separate instance with the first four queries for each method before technical repetitions; every measured repetition creates a new instance. Constructor method order in the new driver uses the explicit R25-constructor|sid|query-position seed. All new processes share the fixed CPU0 affinity and one-thread environment, and all are retained rather than choosing the fastest.

The exploratory prefix is R25-sensitivity-v1. Catalogue seeds are prefix|resample|dataset|rep; queries use prefix|query|dataset|N|rep|k|stratum; method orders use prefix|order|dataset|N|rep|k|stratum|locality. Replicates are 0,1,2. Rows are sampled with replacement, including original duplicates, and reidentified; the same resampling prefix produces dependent nested draws at different N. Query arrays and derived original scores remain in the raw records. The 720 configurations cross two datasets, five sizes, four periods, three k values, two strata and three localities. A configuration contains three sessions; a displayed size/period/locality cell pools 18. Counts and ratios are descriptive; they are not a held-out policy-selection test or a physical population bootstrap.

The complete output is all\_720\_configurations.csv, reconstructed by src/make\_figures.py from analysis summaries that were recomputed from raw arrays together with all five new main figures. Each row retains dataset, size, period, k, stratum, locality, paired ratios and measured counters. JSON analysis retains every session and raw-file binding. Original and supplemental raw per-query latency arrays remain separate. Summed signed changes use sums of session medians; Where reported, P50/P95/P99 are per-repetition quantiles then medians over repetitions; the new analyzer explicitly summarizes P99. Raw maxima remain available and are not renamed percentile values. The frozen source code, rather than an abbreviated prose description, specifies all loop/draw orders.

Direct diagnostics have columns parse, domain, contains, retrieve, build, materialize, instrumented\_full and residual. The first six plus residual equal instrumented\_full for every method/query as integer nanoseconds. Nested checks within build are not also counted in contains. Instrumentation costs are part of these perturbed measurements. No attempt is made to estimate an uninstrumented stage by subtracting these values from separately aggregated primary times. Index-memory reporting counts retained Python column-index objects separately from feature-array bytes; it is not peak RSS. Every dataset/size/seed preparation is shown, not just its minimum.

\section*{S3. Finite-element and bolt-record verification details}

A two-block angular mesh meets the rectangular outer corner exactly. Radial corner spacing is quadratically graded toward the hole, with linear radial midside placement; angular midside nodes are evaluated at intermediate angles using the same boundary-mapped geometry, and only the hole-edge nodes lie on a circular arc. Three systematic verification meshes contain 192, 768 and 3,072 elements (641, 2,433 and 9,473 nodes). Jacobians are checked at integration points and a 3 \ensuremath{\times} 3 reference-boundary grid. The stress comparison uses the same nonsingular hole location (0,a) and 21 points at y/a = 1, 1.1, \ldots{}, 3, rather than a mesh-dependent global maximum. Profile NRMSE is RMS(numerical minus analytical stress) divided by RMS(analytical stress). Predeclared acceptance requires finest-grid hole error and profile NRMSE below 1\%, and medium-to-fine hole-stress and compliance changes below 1\%.

For finite plates, medium-to-fine changes must remain below 2\% for hole stress and 1\% for displacement and strain energy. Reaction forces are K u\ensuremath{-}f on constrained degrees of freedom. The global force residual is normalized by the sum of prescribed nodal-load vector magnitudes; the moment residual is normalized by that scale times the quarter-domain diagonal. Their tolerances are 0.001. The relative difference between strain energy and half external work, and the normalized difference between native and independently reconstructed integration-point stresses, must be below 0.01. The latter checks implementation consistency using the same displacement solution, not independent physical truth. A central-candidate zero-load run uses absolute tolerances, and a half-load run checks half displacement/stress and quarter energy. The accepted matrix contains three exact-reference, eighteen finite-plate and two load-check solves. All inputs, native fields and postprocessed arrays are retained. Computation used Python 3.12.14, NumPy 2.3.5 and SciPy 1.18.1 on a Windows host with two Intel Xeon Gold 6130 CPUs; actual runtime thread utilization was not measured. This case is not a new timing experiment.

The independent displacement reference is Arson \cite{ref21}, printed p. 78 (PDF p. 88), Eq. 3.106. For radius r \ensuremath{\geq} a and polar angle \ensuremath{\theta} measured from the tensile x-axis, set s = r/a, q = a/r, G = E/[2(1+\ensuremath{\nu})] and \ensuremath{\kappa} = (3\ensuremath{-}\ensuremath{\nu})/(1+\ensuremath{\nu}). Equations (S1)--(S3) here use N, mm and MPa consistently. This coefficient is for plane stress; the Itasca example \cite{ref19} is not used as a plane-stress displacement source. Exact outer tractions and symmetry determine the reference problem; analytical displacements are read only after the stored numerical field is loaded and never prescribe its outer boundary.

\begin{equation}
u_r=\frac{a\sigma_\infty}{8G}\left[(\kappa-1)s+2q+2\left(s+(\kappa+1)q-q^3\right)\cos(2\theta)\right]
\tag{S1}\label{eq:supplement-1}
\end{equation}

\begin{equation}
u_\theta=-\frac{a\sigma_\infty}{4G}\left[s+(\kappa-1)q+q^3\right]\sin(2\theta)
\tag{S2}\label{eq:supplement-2}
\end{equation}

\begin{equation}
\begin{aligned}u_x&=u_r\cos\theta-u_\theta\sin\theta,\\u_y&=u_r\sin\theta+u_\theta\cos\theta\end{aligned}
\tag{S3}\label{eq:supplement-3}
\end{equation}

For a node set A, use the unweighted vector norm in Eq. (S4), evaluated once over all nodes and once over nodes satisfying r \ensuremath{\leq} 3a. No displacement offset, rigid-body fit, amplitude adjustment or best-fit rotation is removed. The denominator is nonzero for the declared loaded case. The finest maximum absolute nodal-component error is 2.7890\ensuremath{\times}10\textsuperscript{-11} mm; it is a numerical verification result at \ensuremath{\sigma_{\infty}} = 0.001 MPa, not physical measurement accuracy. Constitutive differentiation uses four steps (0.1,0.01,0.001,0.0001 mm) and compares its plane-stress stresses with the analytical stress field, checking sign, angular orientation and \ensuremath{\kappa} independently of FE agreement.

\begin{equation}
e_A=\left(\frac{\sum_{i\in A}\Vert \mathbf u_i^{FE}-\mathbf u_i^{ref}\Vert _2^2}{\sum_{i\in A}\Vert \mathbf u_i^{ref}\Vert _2^2}\right)^{1/2}
\tag{S4}\label{eq:supplement-4}
\end{equation}

\begingroup
\small
\setlength{\tabcolsep}{4pt}
\begin{longtable}{@{}>{\raggedright\arraybackslash}p{0.166667\dimexpr\textwidth-8\tabcolsep\relax}>{\raggedright\arraybackslash}p{0.166667\dimexpr\textwidth-8\tabcolsep\relax}>{\raggedright\arraybackslash}p{0.333333\dimexpr\textwidth-8\tabcolsep\relax}>{\raggedright\arraybackslash}p{0.333333\dimexpr\textwidth-8\tabcolsep\relax}@{}}
\caption{Independent vector displacement relative L2 errors (fractions, not percent).}\label{tab:supplement-5}\\
\toprule
\textbf{Q8 elements} & \textbf{Nodes} & \textbf{Global relative L2} & \textbf{Near-hole relative L2} \\
\midrule
\endfirsthead
\multicolumn{4}{l}{\small\itshape Table \thetable{} (continued)}\\
\toprule
\textbf{Q8 elements} & \textbf{Nodes} & \textbf{Global relative L2} & \textbf{Near-hole relative L2} \\
\midrule
\endhead
\midrule
\multicolumn{4}{r}{\small\itshape Continued on next page}\\
\endfoot
\bottomrule
\endlastfoot
192 & 641 & 2.6043956e-04 & 7.4706325e-04 \\
768 & 2,433 & 1.9570240e-05 & 5.4948513e-05 \\
3,072 & 9,473 & 1.4139261e-06 & 3.9471690e-06 \\
\end{longtable}
\endgroup

The finite catalogue uses a different load: full L = 200 mm, a = 10 mm, W \ensuremath{\in} \{80,100,120\} mm, t \ensuremath{\in} \{4,6,8\} mm and total full-end force 20 kN. A quarter carries 10 kN; \ensuremath{\bar{u}} is its loaded-edge work-conjugate displacement, and full end-to-end extension is 2\ensuremath{\bar{u}}. The ranking is full net volume, not a continuous optimization. Ligament and thickness lower bounds are negated in the four-coordinate upper-threshold interface. The original stress grid \{100,150,200,250,300,350,400\} MPa crosses displacement \{0.025,0.05,0.075,0.1,0.15,0.2,0.3\} mm with ligament \ensuremath{\geq}20 mm, thickness \ensuremath{\geq}4 mm and k=1. None of these query thresholds is asserted to be a material allowable or code limit.

The 108 new near-attribute positions use each candidate attribute at binary64 nextafter below/equal/above while the other coordinates are permissive. The 4,500 crossed positions use every combination of nine candidates, five stress and five displacement fractions across the medium/fine values, five ligament limits and four thickness limits. Counts include coincident threshold vectors; all 4,608 occurrences are retained, with 4,572 unique vectors. The actual medium/fine confusion matrix includes empty ID 0. Coordinate-envelope ambiguity is a distinct test: its independently combined extrema may not equal either physical mesh response. Geometry-change counts compare selections with and without active geometry limits. Raw fea\_queries.json records each expected, medium, optimistic and pessimistic selection.

For NIST, six original workbooks, 92 specimen identifiers, 4,370 force/displacement pairs and 30 conditions remain unmodified. The article describes 91 tests; the discrepancy is not repaired. Two whole conditions, 22A325T200 and 25A325T400, are quarantined because of identity/force-summary conflicts. Four response values per condition remain UNKNOWN, while known temperature and fixture labels remain usable. The original functional workload (5,460 positions, k=1,2,5) and 333 supported boundary requests are preserved in original\_nist\_case. Original 5,460-position outcomes are 66 answered, 4,306 empty and 1,088 incomplete; 27,300 full reports match. The 333 boundary requests add 1,665 matching reports without replacing the original grid.

The supplemental attainable-capacity queries use k=1,3,5, retain duplicate supported requests and treat answered/empty/incomplete as distinct strata. Counts are 994/300/185 requests and 4,970/1,500/925 full-report comparisons, all matching the literal original-row reference. There are 1,026 unique requests and 50 distinct ordered selected sets. A known failure excludes a condition despite other UNKNOWN fields; an unexcluded rank-relevant UNKNOWN keeps the result incomplete. No imputation, temperature interpolation, cross-fixture extrapolation or statistical resistance inference is used.

\section*{S4. Reproduction, provenance and remaining limits}

Before extraction, run the accompanying standard-library helper from the delivery directory: python verify\_review\_package.py Reviewer\_Computation\_R25.zip. It checks this exact ZIP identity and every manifest entry directly inside the archive without extracting or modifying files. This separate command makes the full-file check repeatable alongside the computational verifier below.

Unzip Reviewer\_Computation\_R25.zip into a local working directory. It contains README.md, requirements-lock.txt, protocol.json, src/, inputs/, results/ and expected/, including inputs/original\_r20\_run/ for the original raw run and the original\_fea\_case/ and original\_nist\_case/ subpackages (the exact manifest is authoritative). The source manifest binds inputs and raw arrays; the archive verification records 14,221 checked files. No credentials, third-party full-text papers or public-upload authorizations are included. Use Python 3.12 with an isolated environment; do not reuse an existing environment whose dependency versions are unknown.

PowerShell commands, from the extracted directory:

py -3.12 -m venv .review-venv

.\textbackslash{}.review-venv\textbackslash{}Scripts\textbackslash{}python.exe -m pip install -r requirements-lock.txt

.\textbackslash{}.review-venv\textbackslash{}Scripts\textbackslash{}python.exe -B -X utf8 src/verify\_package.py --out reproduction-check

The verifier requires an isolated environment and checks module locations, source bindings used by its analyzers and original scientific fields. Complete 14,221-file manifest validation is a separate archive check, not a traversal performed by this computational entry point. It recomputes original statistics and the new three-run/sensitivity summaries, then runs core correctness, the counterexample, Q8 self-tests, two representative FE solves, boundary/displacement postprocessing, NIST state-stratified checks and all five new figures. It does not remeasure all performance calls. For a new timing campaign use src/experiments.py as documented in README.md, preserving every run and the frozen configuration matrix. Timing values naturally depend on the new host; exact equality is expected for the supplied-data statistical recomputation, not future elapsed times.

The author-side fresh environment completed all ten stages in 101.391 s. A separately initiated execution from the actual extracted archive also completed all ten stages, in 102.70 s. These durations describe verification including I/O and checks, not API latency. Both reproduced scientific summary fields. The medium exact-traction and fine central-plate solves reproduced all retained numeric fields in that same environment; other archived FE fields were checked and postprocessed rather than all re-solved. Native arrays, expected outputs, actual commands, exit codes and stage logs are retained. The raw timing archive remains immutable; corrected documentary paths do not replace its numbers.

The ZIP is 195,054,054 bytes; SHA-256 is 70f937f00f0e010555b0b2cc0b0edd805ae7ea85dd6474fbf2945c8bbf6d4480. This identifier names the computation archive, not the accompanying revised manuscript files. A final delivery manifest separately binds those documents. The complete 720-configuration CSV and publication figures are recreated by the supplied figure command. The original 27-page paper and detailed original case materials remain in the computation archive for comparison; the revised main text consolidates rather than silently removes their primary findings.

The archival package is a local reviewer deliverable, not an approved public open-source release. Rights-holder approval is required before public distribution of code and any separately licensed inputs. The Ehlers full-method gap, genuine CRediT confirmation and actual author approval remain unresolved. The current numerical evidence cannot establish nonlinear connection behavior, physical experimental replication, assembly collision freedom, comprehensive design-code compliance or universal retrieval speedup.


\begin{thebibliography}{21}
\urlstyle{rm}
\setlength{\emergencystretch}{2em}

\bibitem{ref1}
Hidayat A, Cheema MA, Lin X, Zhang W, Zhang Y (2022) Continuous monitoring of moving skyline and top-k queries. The VLDB Journal 31(3):459--482. \url{https://doi.org/10.1007/s00778-021-00702-4}

\bibitem{ref2}
Huang W, Li G, Tan K-L, Feng J (2012) Efficient safe-region construction for moving top-K spatial keyword queries. In: Proceedings of the 21st ACM International Conference on Information and Knowledge Management, pp 932--941. \url{https://doi.org/10.1145/2396761.2396879}

\bibitem{ref3}
McClain S, Mutschler-Aldine M, Monaghan C, Chiu D, Sawin J, Jarvis PL (2021) Caching support for range query processing on bitmap indices. In: Proceedings of the 33rd International Conference on Scientific and Statistical Database Management, pp 49--60. \url{https://doi.org/10.1145/3468791.3468800}

\bibitem{ref4}
Rahul S, Gupta P, Janardan R, Rajan KS (2011) Efficient top-k queries for orthogonal ranges. In: WALCOM: Algorithms and Computation, vol 6552, pp 110--121. Springer. \url{https://doi.org/10.1007/978-3-642-19094-0_13}

\bibitem{ref5}
Xie M, Lakshmanan LVS, Wood PT (2013) Efficient top-k query answering using cached views. In: Proceedings of the 16th International Conference on Extending Database Technology, pp 489--500. \url{https://doi.org/10.1145/2452376.2452433}

\bibitem{ref6}
Ehlers C (2015) Top-k semantic caching. Doctoral dissertation, University of Passau. Institutional publication catalogue. \url{https://www.ifis.uni-passau.de/veroeffentlichungen}

\bibitem{ref7}
Keller AM, Basu J (1996) A predicate-based caching scheme for client-server database architectures. The VLDB Journal 5(1):35--47. \url{https://doi.org/10.1007/s007780050014}

\bibitem{ref8}
Dar S, Franklin MJ, J{\'o}nsson B{\TH}, Srivastava D, Tan M (1996) Semantic data caching and replacement. In: Proceedings of the 22nd International Conference on Very Large Data Bases, pp 330--341. \url{https://www.vldb.org/conf/1996/P330.PDF}

\bibitem{ref9}
Rahul S, Janardan R (2014) A general technique for top-k geometric intersection query problems. IEEE Transactions on Knowledge and Data Engineering 26(12):2859--2871. \url{https://doi.org/10.1109/TKDE.2014.2316807}

\bibitem{ref10}
Brooks TF, Pope DS, Marcolini MA (1989) Airfoil self-noise [Data set]. UCI Machine Learning Repository. \url{https://doi.org/10.24432/C5VW2C}

\bibitem{ref11}
Yeh I-C (1998) Concrete compressive strength [Data set]. UCI Machine Learning Repository. \url{https://doi.org/10.24432/C5PK67}

\bibitem{ref12}
Brooks TF, Pope DS, Marcolini MA (1989) Airfoil self-noise and prediction. NASA Reference Publication 1218. NASA. \url{https://ntrs.nasa.gov/citations/19890016302}

\bibitem{ref13}
Yeh I-C (1998) Modeling of strength of high-performance concrete using artificial neural networks. Cement and Concrete Research 28(12):1797--1808. \url{https://doi.org/10.1016/S0008-8846(98)00165-3}

\bibitem{ref14}
Hoefler T, Belli R (2015) Scientific benchmarking of parallel computing systems: Twelve ways to tell the masses when reporting performance results. In: Proceedings of the International Conference for High Performance Computing, Networking, Storage and Analysis, article 73, pp 1--12. \url{https://doi.org/10.1145/2807591.2807644}

\bibitem{ref15}
Davison AC, Hinkley DV (1997) Bootstrap methods and their application. Cambridge University Press. \url{https://doi.org/10.1017/CBO9780511802843}

\bibitem{ref16}
Seif M, Vieira L, Peixoto R (2017) Results from double-shear tests of high-strength structural bolts at elevated temperatures [Data set]. National Institute of Standards and Technology. \url{https://doi.org/10.18434/M33T0T}

\bibitem{ref17}
Peixoto RM, Seif MS, Vieira LCM Jr (2017) Double-shear tests of high-strength structural bolts at elevated temperatures. Fire Safety Journal 94:8--21. \url{https://doi.org/10.1016/j.firesaf.2017.09.003}

\bibitem{ref18}
COMSOL (n.d.) Kirsch Infinite Plate Problem. COMSOL Multiphysics 6.3 Application Library. \url{https://doc.comsol.com/6.3/doc/com.comsol.help.models.sme.kirsch_plate/kirsch_plate.html}. Accessed 27 September 2026.

\bibitem{ref19}
Itasca (2024) Cylindrical Hole in an Infinite Elastic Medium. Itasca Software 9.1 documentation, 3DEC examples. Updated 13 August 2024. \url{https://docs.itascacg.com/itasca910/3dec/docproject/source/examples/CylindricalHoleInAnIEM.html}. Accessed 27 September 2026.

\bibitem{ref20}
Virtanen P, Gommers R, Oliphant TE, et al (2020) SciPy 1.0: fundamental algorithms for scientific computing in Python. Nature Methods 17(3):261--272. \url{https://doi.org/10.1038/s41592-019-0686-2}

\bibitem{ref21}
Arson C (2020) Introduction to Theoretical Geomechanics. Georgia Institute of Technology, lecture notes, p 78, Eq. 3.106. \url{https://emi-poromechanics.github.io/resources/Arson_geomech.pdf}. Accessed 19 September 2026.

\end{thebibliography}
\end{document}